\documentclass[12pt,fleqn]{article}
\usepackage{graphicx,cite}
\usepackage{amsmath,amssymb,booktabs}

\allowdisplaybreaks
\def\theequation{\thesection.\arabic{equation}}
\newcommand{\newsection}[1]{\section{#1}\setcounter{equation}{0}}
\newcommand{\newappendix}[1]{\section*{#1}\setcounter{equation}{0}}

\def\be{\begin{equation}}
\def\ee{\end{equation}}
\def\bea{\begin{eqnarray}}
\def\eea{\end{eqnarray}}
\def\nnb{\nonumber}

\newcommand{\scs}{\scriptscriptstyle}
\newcommand{\f}{\frac}
\newcommand{\fm}[2]{{\textstyle \frac{#1}{#2}}}

\newcommand{\al}{\alpha_{\mathrm s}}
\newcommand{\alt}{\widetilde{\alpha}_{\mathrm s}}

\def\eps{\epsilon}
\def\vareps{\varepsilon}

\newcommand{\loopint}[1]{\int \!\! \frac{d^D #1}{i \left(2\pi\right)^D}\!}
\newcommand{\ESGamma}{S_{\Gamma}}
\newcommand{\hs}[1]{\hspace*{#1 pt}}

\newcommand{\MB}[2]{\hs{-12} \int\limits_{\hs{15}_{ #1 -i \,
\infty}}^{\hs{15}^{ #1 +i\, \infty}} \hs{-15} \frac{d #2}{2\pi i}}
\newcommand{\pFq}[5]{\, \! _{#1} F_{#2}( #3 \, ; \, #4 \, ; \, #5 )}

\begin{document}

\begin{titlepage}
\begin{flushright}
TTK-15-08, QFET-2015-03\\
SI-HEP-2015-03, IFT-1/2015\\
TTP15-008, SFB/CPP-14-122
\end{flushright}

\ \\
\begin{center}
\setlength {\baselineskip}{0.3in} 
{\bf\Large The {\boldmath $(Q_7,Q_{1,2})$} contribution to 
{\boldmath $\bar B\to X_s\gamma$} at
{\boldmath ${\mathcal O}\left(\al^2\right)$}}\\[2cm]
{\large Micha{\l} Czakon$^1$,~ Paul Fiedler$^1$,~ Tobias Huber$^2$,~ Miko{\l}aj Misiak$^3$,\\ 
Thomas Schutzmeier$^4$~ and~ Matthias Steinhauser$^5$}\\[1cm]
\setlength{\baselineskip}{5mm}
$^1$~{\it Institut f\"ur Theoretische Teilchenphysik und Kosmologie, RWTH Aachen University,\\
            D-52056 Aachen, Germany.}\\[2mm]
$^2$~{\it Theoretische Physik 1, Naturwissenschaftlich-Technische Fakult\"at,\\ 
Universit\"at Siegen, Walter-Flex-Stra{\ss}e 3, D-57068 Siegen, Germany.}\\[2mm]
$^3$~{\it Institute of Theoretical Physics, University of Warsaw,\\
         Pasteura 5, PL-02-093 Warsaw, Poland.}\\[2mm]
$^4$~{\it Physics Department, Florida State University, Tallahassee, FL, 32306-4350, USA.}\\[2mm]
$^5$~{\it Institut f\"ur Theoretische Teilchenphysik, 
          Karlsruhe Institute of Technology (KIT),\\
          D-76128 Karlsruhe, Germany.}\\[2cm]
{\bf Abstract}\\[5mm]
\end{center} 
Interference between the photonic dipole operator $Q_7$ and the
current-current operators $Q_{1,2}$ gives one of the most important QCD
corrections to the $\bar B\to X_s\gamma$ decay rate. So far, the
${\mathcal O}\left(\al^2\right)$ part of this correction has been known in the
heavy charm quark limit only ($m_c \gg m_b/2$). Here, we evaluate this part at
$m_c=0$, and use both limits in an updated phenomenological study.  Our
prediction for the CP- and isospin-averaged branching ratio in the
Standard Model reads~ ${\mathcal B}^{\rm SM}_{s\gamma} = (3.36 \pm
0.23) \times 10^{-4}$~ for~ $E_\gamma > 1.6\,$GeV.

\end{titlepage}

\newsection{Introduction \label{sec:intro}}

The inclusive weak radiative decay $\,\bar B\to X_s\gamma\,$ is
known to provide valuable tests of the Standard Model (SM), as well as
constraints on beyond-SM physics. Measurements of its CP- and
isospin-averaged branching ratio ${\mathcal B}_{s\gamma}$ at the
$\Upsilon(4S)$ experiments, namely CLEO~\cite{Chen:2001fj},
Belle~\cite{Abe:2001hk, Limosani:2009qg} and Babar~\cite{Lees:2012ym,
Lees:2012ufa, Lees:2012wg, Aubert:2007my}, contribute to the following world
average\footnote{
The new semi-inclusive measurement by Belle~\cite{Saito:2014das} which
supersedes~\cite{Abe:2001hk} is not yet taken into account in this average.}~\cite{Amhis:2014hma}
\be
{\mathcal B}^{\rm exp}_{s\gamma} = (3.43 \pm 0.21 \pm 0.07) \times 10^{-4}
\ee
for $E_\gamma >  E_0 = 1.6\,$GeV in the $B$-meson rest frame. A
significant suppression of the experimental error is expected once Belle~II
begins collecting data in a few years from now~\cite{Aushev:2010bq,
Abe:2010sj}. 

Let us describe the relation of ${\mathcal B}_{s\gamma}$ to decay rates in an untagged
measurement at $\Upsilon(4S)$.  One begins with the CP-averaged decay rates
\be \label{cpav}
\Gamma_0 = \f{\Gamma(\bar B^0 \to X_s\gamma) + \Gamma(B^0 \to X_{\bar s}\gamma)}{2}, \hspace{1cm}
\Gamma_\pm = \f{\Gamma(B^- \to X_s\gamma) + \Gamma(B^+ \to X_{\bar s}\gamma)}{2}.
\ee
Their isospin average ~$\Gamma = (\Gamma_0+\Gamma_\pm)/2$~ and asymmetry 
~$\Delta_{0\pm} = (\Gamma_0-\Gamma_\pm)/(\Gamma_0+\Gamma_\pm)$~ are related 
to ${\mathcal B}_{s\gamma}$ as follows
\be \label{brwidth}
{\mathcal B}_{s\gamma} = \tau_{B^0} \Gamma \left( \f{1 + r_f r_\tau}{1+r_f} + 
\Delta_{0\pm} \f{1 - r_f r_\tau}{1+r_f}\right). 
\ee
Here, $r_\tau = \tau_{B^+}/\tau_{B^0} = 1.076 \pm
0.004$~\cite{Amhis:2014hma} and $r_f = f^{+-}/f^{00} = 1.059 \pm
0.027$~\cite{Amhis:2014hma} are the measured lifetime and production rate
ratios of the charged and neutral $B$-mesons at $\Upsilon(4S)$.  The term
proportional to $\Delta_{0\pm}$ in Eq.~(\ref{brwidth}) contributes only at a
permille level, which follows from the measured value of $\Delta_{0\pm} =
-0.01 \pm 0.06$ (for $E_\gamma > 1.9\,$GeV)~\cite{Aubert:2007my,Aubert:2005cua,Agashe:2014kda}.  

The final state strangeness in Eq.~(\ref{cpav}) ($-1$ for $X_s$ and $+1$ for
$X_{\bar s}$) as well as the neutral $B$-meson flavours have been specified
upon ignoring effects of the $B^0\bar B^0$ and $K^0\bar K^0$ mixing. Taking
the $K^0\bar K^0$ mixing into account amounts to replacing $X_s$ and $X_{\bar
s}$ by $X_{|s|}$ with an unspecified strangeness sign, which leaves $\Gamma_0$
and $\Gamma_\pm$ invariant. Next, taking the $B^0\bar B^0$ mixing into account
amounts to using in $\Gamma_0$ the time-integrated decay rates of mesons whose
flavour is fixed at the production time. Such a change leaves $\Gamma_0$
practically unaffected because mass eigenstates in the $B^0\bar B^0$ system
are very close to being orthogonal ($|p/q|=1$) and having the same decay width
\cite{Agashe:2014kda}.
%
%
In the following, we shall thus ignore the neutral meson mixing effects.

Theoretical calculations of the $\bar B\to X_s\gamma$ decay rate are based on
the equality
\be
\Gamma(\bar B\to X_s\gamma)_{E_\gamma > E_0} = 
\Gamma(b \to X_s^p \gamma)_{E_\gamma > E_0} + \delta\Gamma_{\rm nonp},
\ee
where the first term on the r.h.s.\ stands for the perturbatively calculable
inclusive decay rate of the $b$ quark into charmless partons $X_s^p =
s, sg, sgg, sq\bar q, \ldots$ and the photon. For appropriately chosen $E_0$,
the second term $\delta\Gamma_{\rm nonp}$ becomes small, and is called a
non-perturbative correction. For $E_0 = 1.6\,$GeV, the uncertainty due to poor
knowledge of $\delta\Gamma_{\rm nonp}$ has been estimated to remain below
$5\%$ of the decay rate~\cite{Benzke:2010js}. The non-perturbative
correction is partly correlated with the isospin asymmetry because
$\delta\Gamma_{\rm nonp}$ depends on whether $\bar B = \bar B^0$ or $\bar B =
B^-$~\cite{Benzke:2010js}.

As far as the perturbative contribution $\Gamma(b \to X_s^p \gamma)$ is
 concerned, its determination with an accuracy significantly better than $5\%$
 is what the ongoing calculations aim at. For this purpose, order
 ${\mathcal O}(\al^2)$ corrections need to be evaluated. Moreover, resummation of
 logarithmically enhanced terms like $\left(\al\ln(M_W^2/m_b^2)\right)^n$ is
 necessary at each order of the usual $\al$-expansion.\footnote{
After the resummation, subsequent ${\mathcal O}(1)$, ${\mathcal O}(\al)$ and ${\mathcal O}(\al^2)$ 
terms in this expansion are called Leading Order (LO), Next-to-Leading Order (NLO) and
Next-to-Next-to-Leading Order (NNLO).}
 Such a resummation is most conveniently performed in the framework of an
 effective theory that arises after decoupling of the electroweak-scale
 degrees of freedom. In the SM, which we restrict to in the present paper, one
 decouples the top quark, the Higgs boson and the gauge bosons $W^\pm$ and
 $Z^0$. Barring higher-order electroweak corrections, all the relevant
 interactions are then described by the following effective Lagrangian:
\be 
{\mathcal L}_{\rm eff} = {\mathcal L}_{\scs {\rm QCD} \times {\rm QED}}(u,d,s,c,b) 
+ \f{4 G_F}{\sqrt{2}} \left[ V^*_{ts} V_{tb} \sum_{i=1}^{8} C_i(\mu) Q_i
                            +  V^*_{us} V_{ub} \sum_{i=1}^{2} C_i(\mu) (Q_i - Q^u_i) \right],
\label{Leff}
\ee
where $G_F$ is the Fermi constant, and $V_{ij}$ are the Cabibbo-Kobayashi-Maskawa (CKM)
matrix elements. The operators $Q^{(u)}_i$ are given by
\bea 
Q^u_1 &=& (\bar{s}_L \gamma_{\mu} T^a u_L) (\bar{u}_L     \gamma^{\mu} T^a b_L),\nnb\\[0mm] 
Q^u_2 &=& (\bar{s}_L \gamma_{\mu}     u_L) (\bar{u}_L     \gamma^{\mu}     b_L),\nnb\\[0mm]
Q_1   &=& (\bar{s}_L \gamma_{\mu} T^a c_L) (\bar{c}_L     \gamma^{\mu} T^a b_L),\nnb\\[0mm]
Q_2   &=& (\bar{s}_L \gamma_{\mu}     c_L) (\bar{c}_L     \gamma^{\mu}     b_L),\nnb\\[0mm]
Q_3   &=& (\bar{s}_L \gamma_{\mu}     b_L) \sum_q (\bar{q}\gamma^{\mu}     q),\nnb\\[0mm]
Q_4   &=& (\bar{s}_L \gamma_{\mu} T^a b_L) \sum_q (\bar{q}\gamma^{\mu} T^a q),\nnb\\[0mm]
Q_5   &=& (\bar{s}_L \gamma_{\mu_1}
                     \gamma_{\mu_2}
                     \gamma_{\mu_3}    b_L)\sum_q (\bar{q} \gamma^{\mu_1} 
                                                           \gamma^{\mu_2}
                                                           \gamma^{\mu_3}     q),\nnb\\[0mm]
Q_6   &=& (\bar{s}_L \gamma_{\mu_1}
                   \gamma_{\mu_2}
                   \gamma_{\mu_3} T^a b_L)\sum_q (\bar{q} \gamma^{\mu_1} 
                                                          \gamma^{\mu_2}
                                                          \gamma^{\mu_3} T^a q),\nnb\\[0mm]
Q_7  &=&  \f{e}{16\pi^2} m_b (\bar{s}_L \sigma^{\mu \nu}     b_R) F_{\mu \nu},\nnb\\[0mm]
Q_8  &=&  \f{g}{16\pi^2} m_b (\bar{s}_L \sigma^{\mu \nu} T^a b_R) G_{\mu \nu}^a,\label{operators}
\eea
where the sums in $Q_{3,\ldots,6}$~ go over all the active flavours $q
= u, d, s, c, b$ in the effective theory.

Decoupling (matching) calculations give us values of the electroweak-scale
Wilson coefficients $C_i(\mu_0)$, where $\mu_0 \sim (M_W,m_t)$.  Next,
renormalization group equations are used to evolve them down to the low-energy
scale, i.e.\ to find $C_i(\mu_b)$, where $\mu_b \sim m_b/2$~ is of order
of the final hadronic state energy in the $\bar B$-meson rest
frame. Determination of the Wilson coefficients $C_{1,\dots,8}(\mu_b)$ up to
${\mathcal O}(\al^2)$ in the SM was completed in 2006~\cite{Bobeth:1999mk,
Misiak:2004ew, Gorbahn:2004my, Gorbahn:2005sa, Czakon:2006ss}. Matching
calculations up to three loops~\cite{Misiak:2004ew} and anomalous dimension
matrices up to four loops~\cite{Czakon:2006ss} were necessary for this 
purpose. The three-loop matching calculation has recently been
extended to the Two-Higgs-Doublet-Model case~\cite{Hermann:2012fc}. 
Most of the final results have been presented for the so-called effective
coefficients
\be \label{eff.coeffs}
C_i^{\rm eff}(\mu) = \left\{ \begin{array}{ll}
C_i(\mu), & \mbox{ for $i = 1,\ldots,6$,} \\[1mm] 
C_7(\mu) + \sum_{j=1}^6 y_j C_j(\mu), & \mbox{ for $i = 7$,} \\[1mm]
C_8(\mu) + \sum_{j=1}^6 z_j C_j(\mu), & \mbox{ for $i = 8$,}
\end{array} \right.
\ee
where the numbers $y_j$ and $z_j$ are such that the LO decay amplitudes for $b
\to s \gamma$ and $b \to sg$ are proportional to the LO terms in $C_7^{\rm
eff}(\mu_b)$ and $C_8^{\rm eff}(\mu_b)$, respectively~\cite{Buras:1994xp}.  In
the $\overline{\rm MS}$ scheme with fully anticommuting $\gamma_5$, one finds
$\vec{y} = (0, 0, -\f{1}{3}, -\f{4}{9}, -\f{20}{3}, -\f{80}{9})$ and $\vec{z}
= (0, 0, 1, -\f{1}{6}, 20, -\f{10}{3})$~\cite{Chetyrkin:1996vx}.

Once the Wilson coefficients $C^{\rm eff}_i(\mu_b)$ have been found up to the
NNLO, one proceeds to evaluating all the on-shell decay amplitudes that
matter at this order for\footnote{
Following the notation of Ref.~\cite{Blokland:2005uk}, we use tilde over G in
the r.h.s.\ of Eq.~(\ref{rate.pole}) to indicate the overall normalization to
$m_{b,\rm pole}^5$.}
\mathindent0cm
\bea 
\Gamma(b \to X_s^p \gamma)_{E_\gamma > E_0} &=& 
\f{G_F^2 \alpha_{\mathrm em} m_{b,\rm pole}^5}{32 \pi^4} \left|V_{ts}^* V_{tb} \right|^2
\sum_{i,j=1}^8 C_i^{\rm eff}(\mu_b) \ C_j^{\rm eff}(\mu_b) \times\nnb\\ 
&\times&
\left[ \widetilde{G}^{(0)}_{ij}(E_0) + \f{\al}{4\pi}\, \widetilde{G}^{(1)}_{ij}(E_0,\mu_b) 
+ \left( \f{\al}{4\pi} \right)^2 \widetilde{G}^{(2)}_{ij}(E_0,\mu_b) + {\mathcal O}(\al^3)\right]
+ \ldots \, , \label{rate.pole}
\eea
\mathindent1cm
where ellipses stand for higher-order electroweak corrections. At the LO, the
symmetric matrix $\widetilde{G}^{(0)}_{ij}$ takes the form
\be \label{gLO}
\widetilde{G}^{(0)}_{ij}(E_0) = \delta_{i7} \delta_{j7} + T^{(0)}_{ij},
\ee
where $T^{(0)}_{ij}$ describe small tree-level contributions to $b \to sq{\bar q}\gamma$ from $Q_{1,2}^u$ and 
$Q_{3,\ldots,6}$~\cite{Kaminski:2012eb,Huber:2014nna}. At the NLO and NNLO, numerically
dominant effects come from $\widetilde{G}^{(n)}_{77}$,
$\widetilde{G}^{(n)}_{17}$ and $\widetilde{G}^{(n)}_{27}$. While
$\widetilde{G}^{(2)}_{77}$ is known in a complete
manner~\cite{Blokland:2005uk, Melnikov:2005bx, Asatrian:2006ph,
Asatrian:2006sm, Asatrian:2006rq}, calculations of $\widetilde{G}^{(2)}_{17}$
and $\widetilde{G}^{(2)}_{27}$ are still in progress. Contributions from
massless and massive fermion loops on the gluon lines have been found in
Refs.~\cite{Bieri:2003ue,Ligeti:1999ea,Boughezal:2007ny}, and served as a
basis for applying the Brodsky-Lepage-Mackenzie (BLM)
approximation~\cite{Brodsky:1982gc}. The remaining (non-BLM) parts
of $\widetilde{G}^{(2)}_{(1,2)7}$ have been known so far in the heavy charm
quark limit only ($m_c \gg m_b/2$)~\cite{Misiak:2006ab,Misiak:2010sk}.

In the present work, we evaluate the full $\widetilde{G}^{(2)}_{(1,2)7}$ for
$m_c=E_0=0$. It is achieved by calculating imaginary parts of several hundreds
four-loop propagator-type diagrams with massive internal lines. Next, both
limits are used to interpolate in $m_c$ those parts of the non-BLM
contributions to $\widetilde{G}^{(2)}_{(1,2)7}$ whose exact $m_c$-dependence
is not yet known. It will give us an estimate of their values at the
measured value of $m_c$, and for non-vanishing $E_0$.

Our current approach differs in several aspects from the one in
Ref.~\cite{Misiak:2006ab} where interpolation in $m_c$ was applied to a
combined non-BLM effect from all the $\widetilde{G}^{(2)}_{ij}$ with $i,j\in
\{1,2,7,8\}$.\footnote{
At the NNLO level, we neglect the small Wilson coefficients
$C_3,\ldots,C_6$, and the CKM-suppressed effects from $Q^u_{1,2}$.}
In the present paper, the only interpolated quantities are the above-mentioned
parts of $\widetilde{G}^{(2)}_{(1,2)7}$. Exact $m_c$-dependence of most of
the other important non-BLM contributions to $\widetilde{G}^{(2)}_{ij}$
is now available thanks to calculations performed in
Refs.~\cite{Asatrian:2006rq, Boughezal:2007ny, Ewerth:2008nv}. Last but not
least, the current analysis includes the previously unknown $m_c$-independent
part of $\widetilde{G}^{(2)}_{78}$~\cite{Asatrian:2010rq}, all the relevant
BLM corrections to $\widetilde{G}^{(2)}_{ij}$ with $i,j\neq
7$~\cite{Ligeti:1999ea,Ferroglia:2010xe,Misiak:2010tk}, tree-level
contributions $T^{(0)}_{ij}$~\cite{Kaminski:2012eb,Huber:2014nna},
four-body NLO corrections~\cite{Huber:2014nna}, as well as the updated
non-perturbative corrections~\cite{Benzke:2010js,Ewerth:2009yr,Alberti:2013kxa}. 
The only contributions to $\widetilde{G}^{(2)}_{ij}$ with $i,j\in
\{1,2,7,8\}$ that remain neglected are the unknown $(n \ge 3)$-body final
state contributions to the non-BLM parts of $\widetilde{G}^{(2)}_{ij}$
with $i,j\neq 7$.

The article is organized as follows. In Section~\ref{sec:zero.mc}, we describe
the calculation of $\widetilde{G}^{(2)}_{(1,2)7}$ for $m_c=E_0=0$.  A new
phenomenological analysis begins in Section~\ref{sec:mcdep} where
$m_c$-dependence of the considered correction is discussed, and the
corresponding uncertainty is estimated. In Section~\ref{sec:pheno}, we
evaluate our current prediction for ${\mathcal B}_{s\gamma}$ in the SM,
which constitutes an update of the one given in Ref.~\cite{Misiak:2006zs}.  We
conclude in Section~\ref{sec:concl}. Appendix~A contains results for all the
massless master integrals that were necessary for the calculation in
Section~\ref{sec:zero.mc}. Several relations to quantities encountered in
Ref.~\cite{Buras:2002tp} are presented in Appendix~B. In Appendix~C, we
collect some of the relevant NLO quantities. Appendix~D contains a list of 
input parameters for our numerical analysis together with a correlation
matrix for a subset of them.

\newsection{Calculation of $\widetilde{G}^{(2)}_{17}$ and $\widetilde{G}^{(2)}_{27}$ for $m_c=E_0=0$ 
\label{sec:zero.mc}}

\subsection{The bare calculation \label{subsec:bare}}

\begin{figure}[t]
\begin{center}
\includegraphics[width=140mm,angle=0]{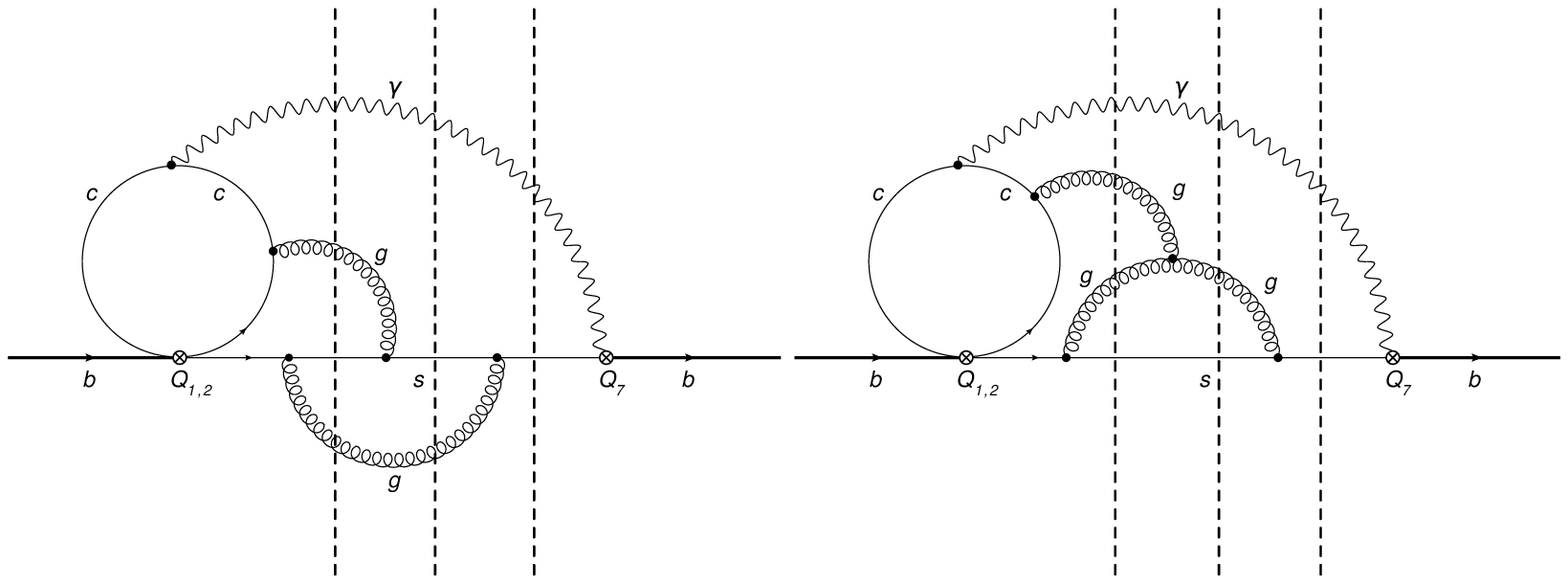}\\[1cm]
\includegraphics[width=140mm,angle=0]{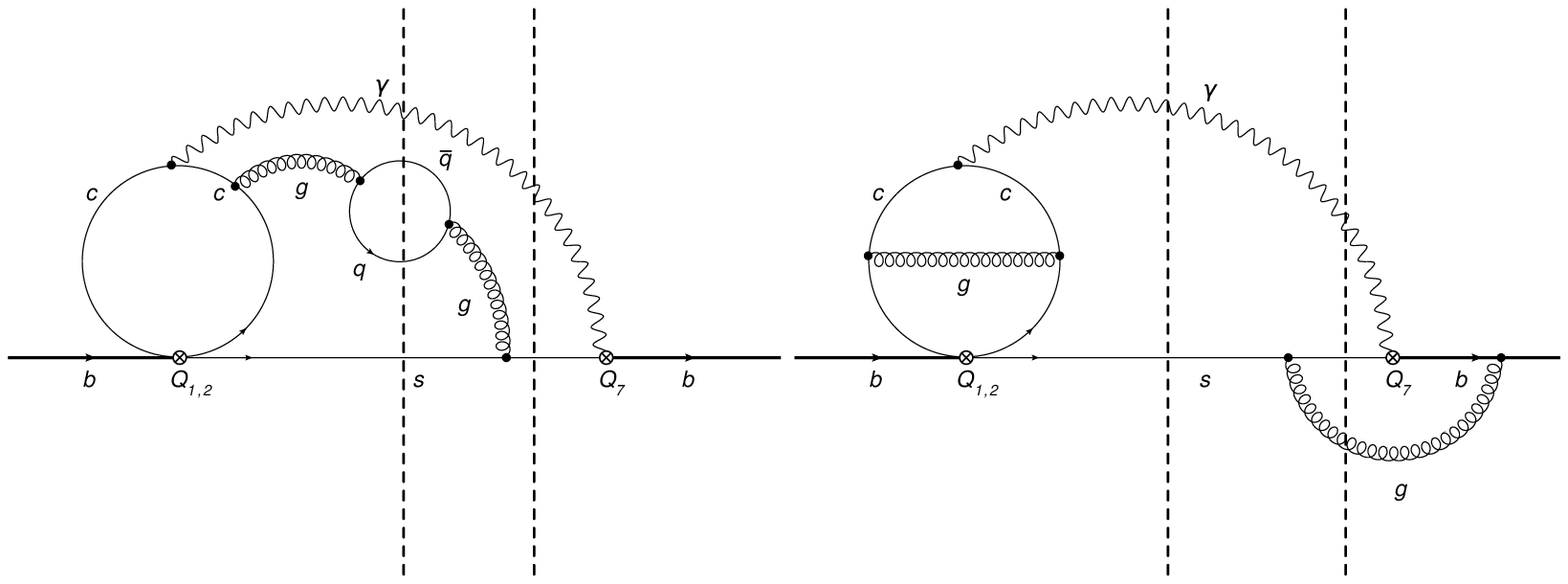}
\caption{\sf Sample diagrams for $\widetilde{G}^{(2)}_{(1,2)7}$
with some of the possible cuts indicated by the dashed lines. \label{fig:diags}}
\end{center}
\end{figure}

Typical diagrams that had to be evaluated for the present project are shown in
Fig.~\ref{fig:diags}. They represent a subset of possible unitarity cut
contributions to the $b$-quark self-energy due to the interference of various
effective operators. At the highest loop level, i.e.\ four-loops, this
interference involves the operators $Q_{1,2}$ and $Q_7$. We need to consider
two-, three- and four-particle cuts. Possible five-particle cuts would
necessarily involve real $c\bar c$ pairs originating from the $Q_{1,2}$
operator vertices, while open charm production is not included in $\bar B \to
X_s \gamma$ by definition. For this reason, we skip the diagrams with
five-particle cuts together with all the diagrams with real $c\bar c$
production or virtual charm loops on the gluon lines. In
Section~\ref{sec:mcdep}, contributions from virtual charm loops on the gluon
lines will be taken over from the $m_c \neq 0$ calculation of
Ref.~\cite{Boughezal:2007ny}, and added to the final result.

For efficiency reasons, we work directly with cut diagrams and employ
the technique first proposed in \cite{Anastasiou:2002yz}. The idea of
the method is to represent cut propagators as
\begin{equation}
-2\pi i\delta(p^2-m^2) = \frac{1}{p^2-m^2+i
  \vareps}-\frac{1}{p^2-m^2-i \vareps} \; .
\end{equation}
As long as we perform only algebraic transformations on the integrands, there
is no difference between the first and second terms on the r.h.s.\ of the
above equation, and it is sufficient to work with one of them only. This is
particularly convenient for the integration-by-parts (IBP) method for
reduction of integrals~\cite{Chetyrkin:1981qh}. The only difference in
such an approach between complete integrals and cut integrals is that a given
integral vanishes if the cut propagator disappears due to cancellation of
numerators with denominators. This fact reduces the number of occurring
integrals in comparison to a computation without cuts.

In practice, the calculation follows the standard procedure. Diagrams are
generated with {\tt DiaGen} \cite{DiaGenIdSolver}, the Dirac algebra is
performed with {\tt FORM}~\cite{Kuipers:2012rf}, and the resulting
scalar integrals are reduced using IBP identities with {\tt IdSolver}
\cite{DiaGenIdSolver}. The main challenge of this calculation begins after
these steps. The amplitudes for the interference contributions are expressed
in terms of a number of master integrals, most of them containing massive
internal $b$-quark lines and a non-trivial phase space integration in
$D=4-2\eps$ spacetime dimensions, with up to four particles in the final
state. A feeling for the size of the problem can be gained from
Tab.~\ref{tab:integrals}.
\begin{table}[t]
\begin{center}
\begin{tabular}{r||cccc}
       & $n_{D}$ & $n_{OS}$ & $n_{\rm eff}$ & $n_{\rm massless}$ \\
\hline\hline
two-particle cuts   & 292 &  92 & 143 &  9 \\
three-particle cuts & 267 &  54 & 110 & 11 \\
four-particle cuts  & 292 &  17 &  37 &  7 \\ \hline
total               & 851 & 163 & 290 & 27 
\end{tabular}
\end{center}
\caption{\sf Number of diagrams $n_{{D}}$, number of massive on-shell master integrals $n_{{OS}}$, number of
effectively computed massive master integrals $n_{\rm eff}$, and number
of massless master integrals $n_{\rm massless}$. The last two columns
are explained in the text.\label{tab:integrals}}
\end{table}

\begin{figure}[t]
\begin{center}
\includegraphics[width=120mm,angle=0]{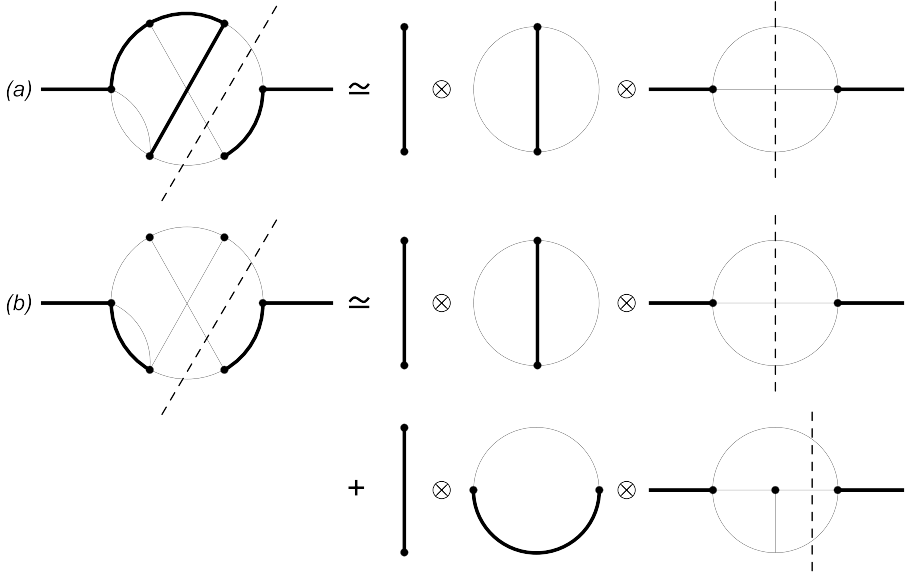}
\caption{\sf Diagrammatic representation of the asymptotic large mass
expansion of two non-planar master integrals.Thick and thin lines 
represent massive and massless propagators, respectively, while dashed 
lines show the unitarity cuts.\label{fig:lme}}
\end{center}
\end{figure}

Having a large number of massive cut integrals, it is
advantageous to devise a strategy to treat them in a uniform manner. It
is clear that purely massless cut integrals are easier to calculate than
massive ones. Therefore, we aim at replacing a calculation of massive 
propagator integrals by a calculation of massless ones. This can be
achieved by extending the integral definitions. We assume, namely, that
the external momentum squared $p_b^2$ is a free parameter, and
treat coefficients ${\mathcal I}_i$ in the $\eps$-expansion of the master
integrals as functions of a single dimensionless variable $x =
p_b^2/m_b^2$. IBP identities give us differential equations
\begin{equation} \label{diff.eq}
\frac{d}{d x} {\mathcal I}_i(x) = \sum_j {\mathcal J}_{ij}(x) {\mathcal I}_j(x) \; ,
\end{equation}
with ${\mathcal J}_{ij}(x)$ being certain rational functions of $x$. Boundary
conditions for these equations in the vicinity of $x=0$ are given by
asymptotic large-mass expansions, i.e.\ by power-log series in $x$. A few
leading terms in the series for each ${\mathcal I}_i$ can be found by
calculating products of massive tadpole integrals up to three loops and
massless propagator ones up to four loops, as illustrated in
Fig.~\ref{fig:lme}. Next, higher-order terms can be determined from the
differential equations themselves by substituting ${\mathcal I}_i$ in terms of
power-log series in $x$. For our application it turns out that around 50 terms
are sufficient to obtain the desired accuracy. This gives us high-precision
boundary conditions at small but non-vanishing $x$ for solving the
differential equations~(\ref{diff.eq}) numerically.

On the way from the vicinity of $x=0$ to the physical point at $x=1$, one
often encounters spurious singularities on the real axis. To bypass them, the
differential equations are solved along ellipses in the complex
$x$ plane. Several such ellipses are usually considered to test whether the
numerical solution is stable.

\begin{figure}[t]
\includegraphics[width=80mm,angle=0]{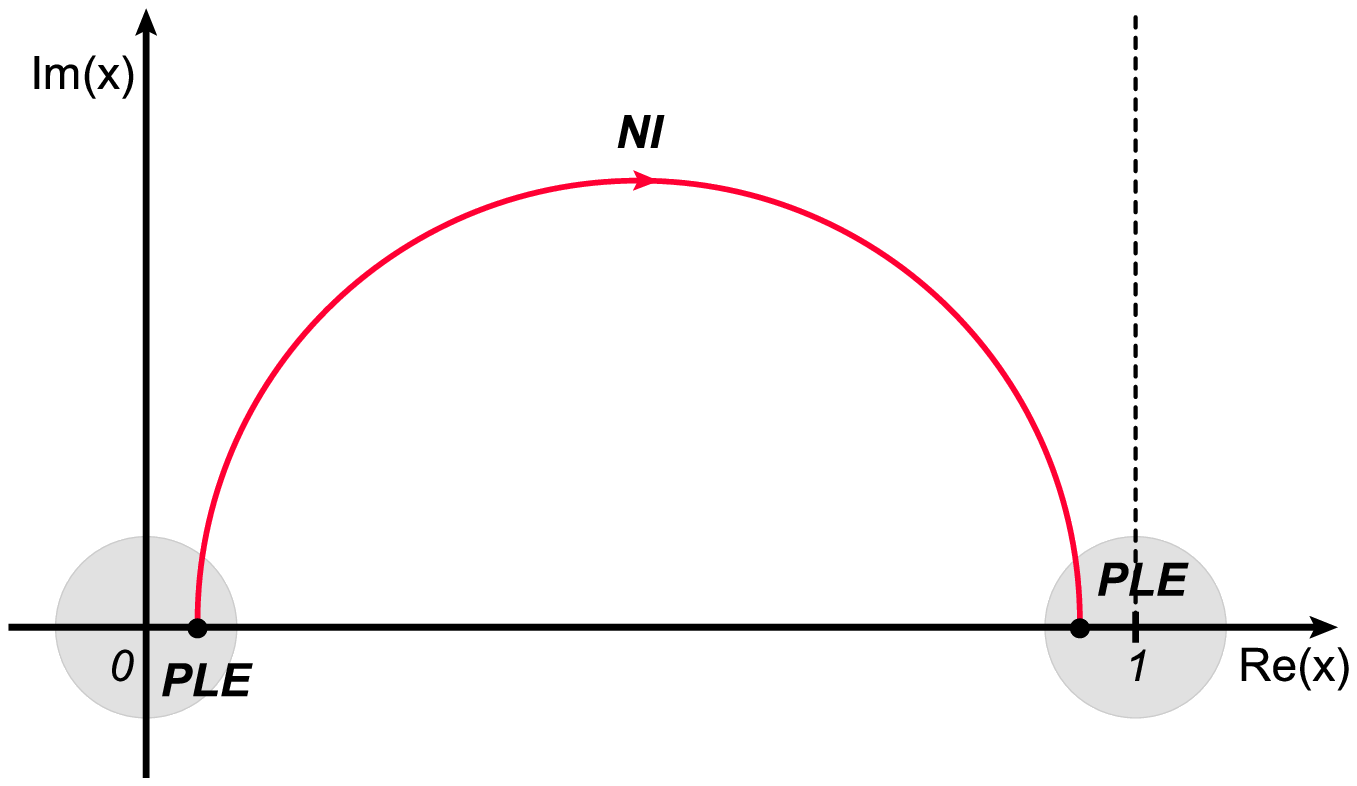}\\[-43mm]
\hspace*{9cm}
\includegraphics[width=80mm,angle=0]{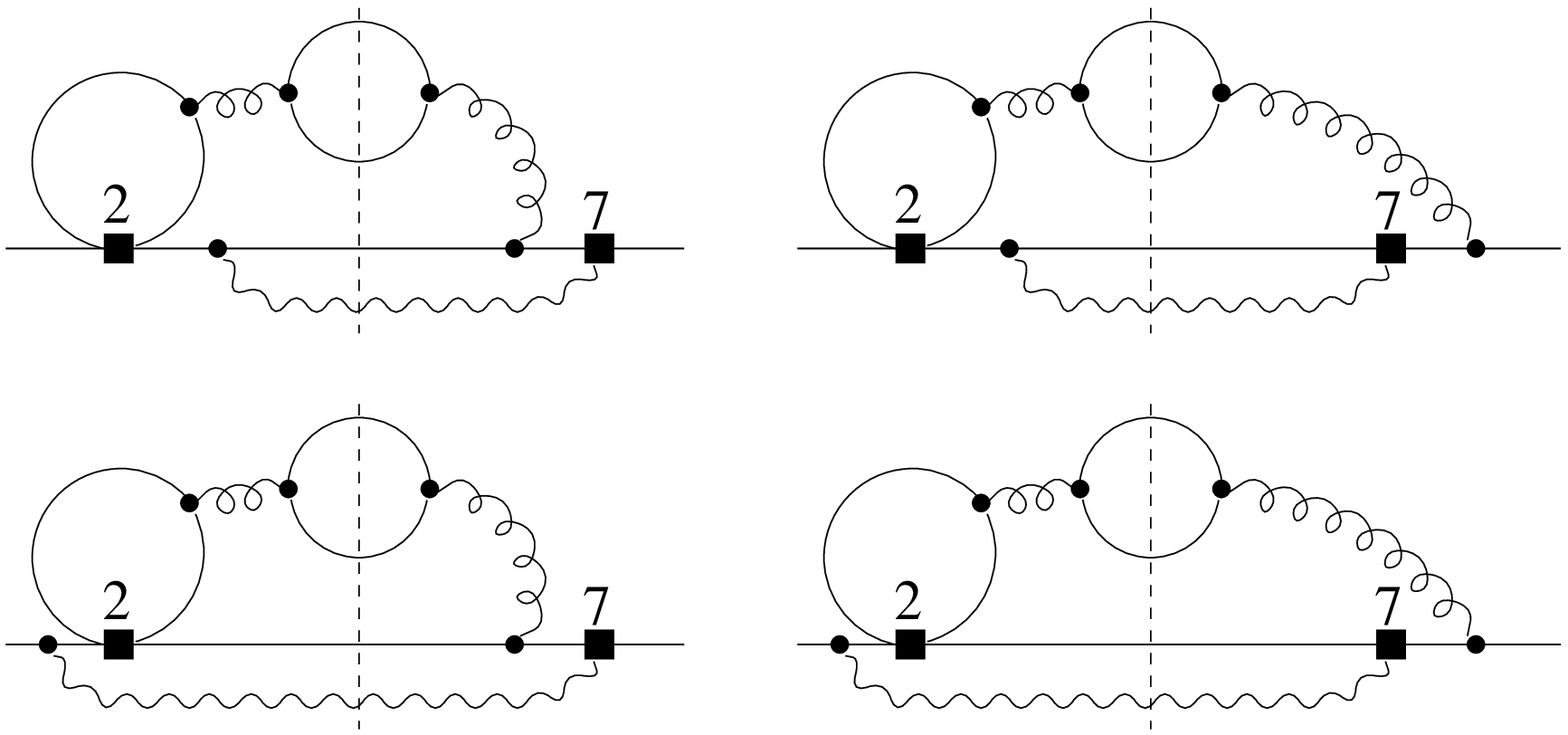}
\begin{center}
\caption{\sf Left (a): Integration contour in the complex $x$ plane.  The
numerical integration (NI) is performed between the regions close to $x=0$ and
$x=1$ that are accessible by power-log expansions (PLE).~ Right (b): Diagrams that
give the terms marked with $\kappa$ in Eq.~(\ref{bare.nnlo}).\label{fig:ab}}
\end{center}
\end{figure}

Naively, one might think that as long as there are no infinities at $x=1$, the
numerical solution could be continued up to that point. However, there is an
essential singularity there, and the integrals behave as $(1-x)^n \ln^m
(1-x)$, with $n,m > 0$ being some positive powers. Due to such a behaviour,
the numerical solution has poor convergence, as the algorithms assume locally
polynomial behaviour of the considered functions.  In order to overcome this
problem, we perform another power-log expansion around $x=1$, and match it
onto the numerical result.  To determine the maximal power of the
logarithms, we begin with observing that the highest poles in the cut diagrams
could potentially be of order $1/\epsilon^6$, due to the presence of collinear
and soft divergences. The coefficient of the leading singularity contains no~
$\ln(1-x)$ because logarithms are generated by expanding expressions of the
form $(1-x)^{a \epsilon}/\epsilon^6$ (with $a$ being some constant) in the
framework of expansion by regions. Thus, finite parts of the master integral
expansions may only contain $\ln^6(1-x)$. Higher powers may be needed due to
the presence of spurious singularities, i.e.\ poles in the coefficients at the
master integrals in the physical amplitude. In practice, we have used an
ansatz with logarithm powers up to fifteen. Our numerical matching has shown
that such high powers never occur in the considered problem, i.e.\ the respective
expansion coefficients are consistent with zero to very high numerical
precision. Using the matched series, we finally obtain the required values
of the original master integrals at $x=1$. The solution procedure is
schematically represented in Fig.~\ref{fig:ab}a.

Since the master integrals are considered for $x\neq 1$, their overall
number $n_{\rm eff}$ is larger than it would be for $x=1$, i.e.\ $n_{\rm eff} >
n_{OS}$. However, the massless integrals that are necessary to determine the
boundary conditions near $x=0$ are not only simpler, but also their number
$n_{\rm massless}$ is much smaller than $n_{OS}$, as seen in
Tab.~\ref{tab:integrals}. All the massless integrals that we had to consider
are depicted in Appendix~A, in Fig.~\ref{fig:4PCutMasters} and
Tab.~\ref{tab:masters}.

Using the above method, we have obtained the following bare NNLO results for
the considered interferences in the Feynman-'t~Hooft gauge:
\bea
\widetilde{G}_{17}^{(2) \rm bare} &=& -\f{1}{6} \widetilde{G}_{27}^{(2) \rm bare} +
\f{80}{81\,\eps^2} + \f{1592 + 54 \pi^2}{243\, \eps} + 42.0026519628,\nnb\\[2mm]
\widetilde{G}_{27}^{(2) \rm bare} &=&  -\f{4}{3\,\eps^3} -\f{30332 + 432 \pi^2}{2187\,\eps^2}
- \f{67.66077706444119}{\eps} + 44.5070537274\nnb\\[2mm]
&+& \kappa\, n_l \left( \f{32}{729\,\eps} + 0.6520676315 \right)
+ n_l \left( \f{352}{729\,\eps^2} + \f{11624}{2187\,\eps} 
+ \f{228656}{6561} - \f{188}{243} \pi^2 \right)\nnb\\[2mm]
&+& n_b \left( \f{352}{729\,\eps^2} + \f{5.17409838118169}{\eps} 
+ 15.1790288135\right) + {\mathcal O}(\eps).\label{bare.nnlo}
\eea
Here, $n_l$ and $n_b$ denote numbers of massless and massive ($m=m_b$) quark
flavours, while $\kappa=1$ marks contributions from the diagrams in
Fig.~\ref{fig:ab}b describing interferences involving four-body $sq\bar
q\gamma$ final states and no $c\bar c\gamma$ couplings. The terms
proportional to $n_l$ and $n_b$ but not marked by $\kappa$ reproduce (after
renormalization) the $m_c \to 0$ limits of what is already known for non-zero
$m_c$~\cite{Bieri:2003ue,Ligeti:1999ea,Boughezal:2007ny}.  For compactness,
all the results in this subsection are given for $\mu^2=e^\gamma
m_b^2/(4\pi)$, where $\gamma$ is the Euler-Mascheroni constant.

Some of the numbers in Eq.~(\ref{bare.nnlo}) have been given in an exact form
even though our calculation of the master integrals at $x=1$ is purely
numerical. However, the accuracy is very high (to around 14 decimals), so
identification of simple rationals is possible. Moreover,
renormalization gives us relations to lower-order results where more terms are
known in an exact manner (see below). For the $n_l$-term, after verifying
numerical agreement with Refs.~\cite{Bieri:2003ue,Misiak:2010tk}, we have made
use of the available exact expressions.\footnote{
In particular, for the function given in Eq.~(13) of Ref.~\cite{Misiak:2010tk},
we have~ $\lim_{m_c\to 0} h_{27}^{(2)}(\delta = 1) = \f{41}{27} - \f{2}{9} \pi^2$.}
Several other numbers in this subsection that have been retained in a decimal
form can actually be related to quantities encountered in
Ref.~\cite{Buras:2002tp}, as described in Appendix~B.

Let us now list all the lower-order bare contributions that are needed for
renormalization. For this purpose, it is convenient to express
Eq.~(\ref{rate.pole}) in terms of $C_i$ rather than $C_i^{\rm eff}$, and
denote the corresponding interference terms by $\hat{G}_{ij}^{(n)}$ rather
than $\widetilde{G}_{ij}^{(n)}$. All the necessary $\hat{G}_{i7}^{(0)}$ and
$\hat{G}_{i7}^{(1)\rm bare}$ read\footnote{
$\hat{G}_{i7}$ differ from $\widetilde{G}_{i7}$ only for $i=3,4,5,6$.}
\bea
\hat{G}_{77}^{(0)} &=& \f{\Gamma(2-\eps)\, e^{\gamma\eps}}{\Gamma(2-2\eps)},\nnb\\[2mm]
\hat{G}_{47}^{(0)} &=& \f43 \hat{G}_{37}^{(0)} ~=~ -\f49\,
        \Gamma(1+\eps)\; e^{\gamma\eps}\; \hat{G}_{77}^{(0)},\nnb\\[2mm]
\hat{G}_{67}^{(0)} &=& \f43 \hat{G}_{57}^{(0)} ~=~ 
        4 \left( 5 - 3\, \eps - \, \eps^2 \right)\, \hat{G}_{47}^{(0)},\nnb\\[2mm]
\hat{G}_{27}^{(1) \rm bare} &=& -6\, \hat{G}_{17}^{(1) \rm bare} ~=~ 
       -\f{92}{81\,\eps} -\f{1978}{243} +\f{777\pi^2-27185}{729}\,\eps + {\mathcal O}(\eps^2),\nnb\\[2mm]
\hat{G}_{47}^{(1) \rm bare} &=& 
       \f{16}{3\, \eps^2} + \f{3674}{243\, \eps} + 43.76456245573869 + 94.9884724116\, \eps \nnb\\[2mm]
       &+& \kappa\, n_l \left( -\f{16}{243} + \f{44 \pi^2 - 612}{243}\, \eps \right) 
       + n_l \left( \f{16}{81\, \eps} - \f{4}{243} + \f{264 \pi^2 - 2186}{729}\, \eps \right)\nnb\\[2mm]
       &+& n_b \left( \f{16}{81\, \eps} + 0.04680853247986 + 0.3194493123\, \eps \right)
       + {\mathcal O}(\eps^2),\nnb\\[2mm]
\hat{G}_{77}^{(1) \rm bare} &=& \f{4}{3\, \eps} + \f{124}{9} - \f{16}{9} \pi^2 + 
       \left( \f{212}{3} - \f{58}{9} \pi^2 - \f{64}{3} \zeta_3 \right) \eps + {\mathcal O}(\eps^2),\nnb\\[2mm]
\hat{G}_{78}^{(1) \rm bare} &=& \f{16}{9\, \eps} + \f{280}{27} - \f{16}{27} \pi^2 + 
       \left( \f{382}{9} - \f{16}{9} \pi^2 - \f{160}{9} \zeta_3 \right) \eps + {\mathcal O}(\eps^2),\nnb\\[2mm]
\hat{G}_{7(12)}^{(1) \rm bare} &=& -6\, \hat{G}_{7(11)}^{(1) \rm bare} ~=~
       \f{2096}{81} + \f{39832}{243}\, \eps + {\mathcal O}(\eps^2).\label{lower.bare}
\eea
The last line of the above equation describes contributions from the so-called evanescent operators
that vanish in four spacetime dimensions 
\bea
Q_{11} &=& (\bar{s}_L \gamma_{\mu_1}
                      \gamma_{\mu_2}
                      \gamma_{\mu_3} T^a c_L)(\bar{c}_L \gamma^{\mu_1}
                                                        \gamma^{\mu_2}
                                                        \gamma^{\mu_3} T^a b_L)
-16 Q_1,\nnb\\[1mm]
Q_{12} &=& (\bar{s}_L \gamma_{\mu_1}
                      \gamma_{\mu_2}
                      \gamma_{\mu_3}     c_L)(\bar{c}_L \gamma^{\mu_1}
                                                        \gamma^{\mu_2}
                                                        \gamma^{\mu_3}     b_L)
-16 Q_2.
\eea
In $\hat{G}_{(1,2)7}^{(1) \rm bare}$, the three-particle-cut contributions alone 
($b \to s \gamma g$) read 
\be
\hat{G}_{27}^{(1)3P} = -6\, \hat{G}_{17}^{(1)3P} = 
         -\f{4}{27} - \f{106}{81}\,\eps + {\mathcal O}(\eps^2).
\ee
In addition, several interferences need to be calculated with the
$b$-quark propagators squared, to account for the renormalization of $m_b$. 
We find
\bea
\hat{G}_{27}^{(1)m} &=& -6\, \hat{G}_{17}^{(1)m}
~=~ -\f{1}{3\,\eps^2} -\f{21+4\pi^2}{81\,\eps} +\f{1085}{81} -\f{161}{972}\pi^2 -\f{40}{27}\zeta_3\nnb\\[1mm]
&+& \left( \f{59071}{486} - \f{1645}{2916} \pi^2 - \f{65}{81} \zeta_3 - \f{7}{81} \pi^4 \right) \eps
    + {\mathcal O}(\eps^2),\nnb\\[2mm]
\hat{G}_{47}^{(0)m} &=& \f{4}{3 \eps} + 2 + \f{50 - 2 \pi^2}{9}\,\eps 
     + \f{94 - 3 \pi^2 - 32 \zeta_3}{9}\, \eps^2 + {\mathcal O}(\eps^3). \label{Gm}
\eea
Our conventions for their global normalization will become clear through the way they enter the 
renormalized NNLO expression in Eq.~(\ref{renor.main}) below.

Some of the diagrams with $Q_4$ insertions contain $b$-quark tadpoles that are
the only source of $1/\eps^2$ terms in $\hat{G}_{47}^{(1) \rm bare}$, and
$1/\eps$ terms in $\hat{G}_{47}^{(0)m}$. Such divergences are actually
necessary to renormalize the $1/\eps^3$ poles in Eq.~(\ref{bare.nnlo}). These
tadpole diagrams have been skipped in the NLO calculation of
Ref.~\cite{Buras:2002tp} because they give no contribution to the renormalized
$\hat{G}_{47}^{(1)}$, i.e.\ they cancel out after renormalization of $m_b$.

Among all the bare interferences given in this section, not only the NNLO ones are
entirely new, but also $\hat{G}_{7(12)}^{(1) \rm bare}$,
$\hat{G}_{27}^{(1)m}$ and $\hat{G}_{47}^{(0)m}$. The remaining LO and NLO results
are extensions of the known ones by another power of $\epsilon$, as necessary
for the current calculation.\footnote{
Exceptions are $\hat{G}_{77}^{(0) \rm bare}$ $\hat{G}_{77}^{(1) \rm bare}$ and
$\hat{G}_{78}^{(1) \rm bare}$, for which sufficiently many terms in the
$\epsilon$ expansions have been already found in
Refs.~\cite{Blokland:2005uk,Asatrian:2006ph,Asatrian:2010rq}. Our results
agree with theirs, barring different conventions for the global $1 + {\mathcal
O}(\eps)$ normalization factor (see the end of subsection~\ref{subsec:renor}).}

\subsection{Renormalization \label{subsec:renor}}

Our results in the previous subsection contain no loop corrections on
external legs in the interfered amplitudes. Such corrections are taken into
account below, with the help of on-shell renormalization constants for the
$b$-quark, $s$-quark and gluon fields
\bea
Z_b^{\rm OS} &=& 1 - \f43\,\alt\,s^\eps\,e^{\gamma\eps}\,\Gamma(\eps)\, 
\f{3 - 2\eps}{1 - 2\eps} + {\mathcal O}(\alt^2),\nnb\\
Z_s^{\rm OS} &=& 1 + {\mathcal O}(\alt^2),\nnb\\
Z_G^{\rm OS} &=& 1 - \f23\,n_b\,\alt\,s^\eps\,e^{\gamma\eps}\,\Gamma(\eps)\, + {\mathcal O}(\alt^2),
\label{OS.const}
\eea
where $\alt = \f{\al}{4\pi} = \f{g_s^2}{16\pi^2}$ and $s = \f{4\pi\mu^2}{m_b^2}e^{-\gamma}$.
The QCD coupling $g_s$ and the Wilson coefficients $C_i$ are renormalized in
the $\overline{\rm MS}$ scheme: $g_s^{\rm bare} = \bar{Z}_g g_s$, and $C_i^{\rm bare} =
\sum_j C_j \bar{Z}_{ji}$. The corresponding ${\rm MS}$ renormalization constants
can be taken over from the literature (see, e.g., Refs.~\cite{Gorbahn:2004my,Czakon:2006ss})
\mathindent0cm
\bea
\begin{array}{rclrcl}
Z_g &=& 1 + \f{\alt}{\eps} \left(-\f{11}{2} + \f{f}{3} \right) + {\mathcal O}(\alt^2),~~~~ &
Z_{77} &=& 1 + \f{16\, \alt}{3\, \eps} + {\mathcal O}(\alt^2),\\[2mm]
Z_{11} &=& 1 - \f{2\, \alt}{\eps} + {\mathcal O}(\alt^2), &
Z_{21} &=& \f{6\, \alt}{\eps} + {\mathcal O}(\alt^2),\\[2mm]
Z_{12} &=& \f{4\, \alt}{3\, \eps} + {\mathcal O}(\alt^2), &
Z_{22} &=& 1 + {\mathcal O}(\alt^2),\\[2mm]
Z_{13} &=& \alt^2 \left( \f{10}{81\,\eps^2} -\f{353}{243\,\eps}\right) + {\mathcal O}(\alt^3), &
Z_{23} &=& \alt^2 \left(-\f{20}{27\,\eps^2} -\f{104}{81\,\eps}\right) + {\mathcal O}(\alt^3),\\[2mm]
Z_{14} &=& -\f{1}{6} Z_{24} + \alt^2 \left( \f{1}{2\eps^2} 
           - \f{11}{12\, \eps} \right), & 
Z_{24} &=& \f{2\, \alt}{3\, \eps} + \alt^2 \left( \f{-188+12f}{27\, \eps^2} 
           + \f{338}{81\, \eps} \right) + {\mathcal O}(\alt^3),\nnb\\[2mm]
Z_{15} &=& \alt^2 \left(-\f{1}{81\, \eps^2} +\f{67}{486\,\eps}\right) + {\mathcal O}(\alt^3), &
Z_{25} &=& \alt^2 \left( \f{2}{27\, \eps^2} +\f{14}{81\, \eps}\right) + {\mathcal O}(\alt^3),\\[2mm]
Z_{16} &=& \alt^2 \left(-\f{5}{216\,\eps^2} -\f{35}{648\,\eps}\right) + {\mathcal O}(\alt^3), &
Z_{26} &=& \alt^2 \left( \f{5}{36\, \eps^2} +\f{35}{108\,\eps}\right) + {\mathcal O}(\alt^3),\\[2mm]
Z_{17} &=& -\f{1}{6} Z_{27} + \alt^2 \left( \f{22}{81\,\eps^2} - \f{332}{243\, \eps} \right), &
Z_{27} &=& \f{116\, \alt}{81\, \eps} + \alt^2 \left( \f{-3556+744f}{2187\, \eps^2} 
           + \f{13610-44f}{2187\, \eps} \right) + {\mathcal O}(\alt^3),\\[2mm]
Z_{18} &=& \f{167\, \alt}{648\, \eps} + {\mathcal O}(\alt^2), &
Z_{28} &=& \f{19\, \alt}{27\, \eps} + {\mathcal O}(\alt^2),\\[2mm]
Z_{1(11)} &=& \f{5\, \alt}{12\, \eps} + {\mathcal O}(\alt^2), &
Z_{2(11)} &=& \f{\alt}{\eps} + {\mathcal O}(\alt^2),\\[2mm]
Z_{1(12)} &=& \f{2\, \alt}{9\, \eps} + {\mathcal O}(\alt^2), &
Z_{2(12)} &=& {\mathcal O}(\alt^2),\\[2mm]
\end{array}\nnb\\[-16mm]\nnb
\eea\bea \label{renconst} \eea
where ~$f=n_l+n_b$~ here, as we have skipped all the charm loops on the gluon
lines.  For the $b$-quark mass renormalization, we use the on-shell scheme
everywhere ($Z_m^{\rm OS} = Z_b^{\rm OS} + {\mathcal O}(\alt^2)$), to get the overall $m_{b,\rm
pole}^5$ in Eq.~(\ref{rate.pole}).

With all the necessary ingredients at hand, we can now write an explicit
formula for the renormalized interference terms up to the NNLO ($i=1,2$)\footnote{
Obviously, the renormalized $\widetilde{G}_{i7}^{(n)}$ remain unchanged after
replacing~ $\bar{Z}_g \to Z_g$,~ $\bar{Z}_{ij} \to Z_{ij}$~
and~ $s \to \mu^2/m_b^2$~ on the r.h.s.\ of Eq.~(\ref{renor.main}) and inside
the on-shell constants (\ref{OS.const}).}
\bea
\alt\, \widetilde{G}_{i7}^{(1)} + \alt^2\, \widetilde{G}_{i7}^{(2)} &=& 
Z_b^{\rm OS}\, Z_m^{\rm OS}\, \bar{Z}_{77} \left\{ \alt^2\, s^{3\eps}\, \widetilde{G}_{i7}^{(2) \rm bare} 
+ (Z_m^{\rm OS}-1)\, s^\eps \left[ \bar{Z}_{i4}\, \hat{G}_{47}^{(0)m}  
+ \alt\, s^\eps\, \hat{G}_{i7}^{(1)m} \right]
\right. \nnb\\[2mm] &+& \left. 
\alt\, (Z_G^{\rm OS}-1)\, s^{2\eps}\, \hat{G}_{i7}^{(1)3P} +
\bar{Z}_{i7}\, Z_m^{\rm OS} \left[ \hat{G}_{77}^{(0)} + \alt\, s^{\eps}\, \hat{G}_{77}^{(1)\rm bare} \right]
+ \alt\, \bar{Z}_{i8}\, s^\eps\, \hat{G}_{78}^{(1)\rm bare} 
\right. \nnb\\[2mm] &+& \left. 
\sum_{j=1,\ldots,6,11,12} \bar{Z}_{ij}\, s^\eps \left[ \hat{G}_{j7}^{(0)} 
+ \alt\, s^\eps\, \bar{Z}_g^2\, \hat{G}_{j7}^{(1)\rm bare} \right] \right\}~
+~ {\mathcal O}(\alt^3),
\label{renor.main}
\eea
\mathindent1cm
where $\hat{G}_{j7}^{(0)} =0$ for $j=1,2,11,12$.  Once the above expression is
expanded in $\alt$, and ${\mathcal O}(\alt^3)$ terms are neglected, all the
$1/\eps^n$ poles cancel out as they should. Our final renormalized results at
$E_0=m_c=0$ read
\bea
\widetilde{G}_{27}^{(1)} &=& -6\, \widetilde{G}_{17}^{(1)} ~=~ 
-\f{1702}{243} \;-\; \f{416}{81} \ln\f{\mu}{m_b},\nnb\\[2mm]
\widetilde{G}_{17}^{(2)} &=& -\f16 \widetilde{G}_{27}^{(2)} \;+\; \f{136}{27} \ln^2\f{\mu}{m_b}
\;+\; \f{94+8\pi^2}{9} \ln\f{\mu}{m_b} \;+\; 22.6049613485,\nnb\\[2mm]
\widetilde{G}_{27}^{(2)} &=& \left( \f{11792}{729} \;+\; \f{800}{243}\,(n_l+n_b)\right) \ln^2\f{\mu}{m_b}
\;+\; \left( 1.0460332197 \;+\; \f{64}{729}\,\kappa\, n_l \right.\nnb\\[2mm]
&+& \left. \f{2368}{243}\, n_l + 9.6604967166\, n_b \right) \ln\f{\mu}{m_b}
\;-\; 14.0663747289 \;+\; 0.1644478609\,\kappa\, n_l\nnb\\[2mm]
&+& \left( \f{54170}{6561} \;+\; \f{92}{729} \pi^2 \right)\, n_l \;-\; 1.8324081161\,n_b. 
\label{renormalized}
\eea
They are, of course, insensitive to conventions for the global $1 + {\mathcal
O}(\eps)$ normalization factor in Eqs.~(\ref{bare.nnlo})--(\ref{Gm}), so long
as it is the same in all these equations. In particular, it does not matter
that our $\hat{G}^{(0)}_{77}$ differs from the one in Ref.~\cite{
Blokland:2005uk} by an overall factor of~ $\Gamma(1+\eps)\,e^{\gamma\eps}$.

As already mentioned, the $n_l$ terms not marked by $\kappa$ in
Eq.~(\ref{renormalized}) agree with the previous calculations where both
$m_c\neq 0$ and $m_c=0$ were considered. In the case of the $n_b$ terms, the
current result extends the published fit (Eq.~(3.3) of
Ref.~\cite{Boughezal:2007ny}) down to $m_c=0$. All the remaining terms are
entirely new.

\newsection{Impact of the NNLO corrections to $(Q_7,Q_{1,2})$ interferences on the branching ratio
\label{sec:mcdep}}

In the description of our phenomenological analysis, we shall strictly follow the 
notation of Ref.~\cite{Misiak:2006ab}, where the relevant perturbative quantity
\be \label{PE0}
P(E_0) ~=~ \sum_{i,j=1}^{8} C_i^{\rm eff}(\mu_b)\; C_j^{\rm eff}(\mu_b)\; K_{ij}(E_0,\mu_b),
\ee
has been defined through
\be \label{pert.ratio}
\f{\Gamma[ b \to X^p_s \gamma]_{E_{\gamma} > E_0}}{
|V_{cb}/V_{ub}|^2 \; \Gamma[ b \to X^p_u e \bar{\nu}]} ~=~ 
\left| \f{ V^*_{ts} V_{tb}}{V_{cb}} \right|^2 
\f{6 \alpha_{\rm em}}{\pi} \; P(E_0).
\ee
The relation between $\widetilde{G}_{i7}^{(n)}$ for~ $i=1,2$~ and $K_{i7} = 
\alt K_{i7}^{(1)} + \alt^2 K_{i7}^{(2)} + {\mathcal O}(\alt^3)$ is thus very simple
\be \label{G2K}
\alt K_{i7}^{(1)} + \alt^2 K_{i7}^{(2)} + {\mathcal O}(\alt^3) ~=~
\f{\alt\, \widetilde{G}_{i7}^{(1)} + \alt^2\, \widetilde{G}_{i7}^{(2)} + {\mathcal O}(\alt^3)}{
1 \;+\; \alt (50 - 8\pi^2)/3 \;+\; {\mathcal O}(\alt^2)}\;,
\ee
where the denominator comes from the NLO correction to the semileptonic
$b \to X^p_u e \bar{\nu}$ decay rate.

In the following, we shall write expressions for $K_{i7}^{(2)}$ that are valid
for arbitrary $m_c$ and $E_0$ but incorporate information from our
calculation in the previous section, where $E_0=m_c=0$ has been assumed. 
For this purpose, four functions
\bea
f_{\scs NLO}(z,\delta) &=& {\rm Re}\,r_2^{(1)}(z) ~+~ 2 \phi^{(1)}_{27}(z,\delta),\nnb\\[1mm]
f_q(z,\delta) &=& {\rm Re}\,r_2^{(2)}(z) ~-~ \fm{4}{3} h^{(2)}_{27}(z,\delta),\nnb\\[1mm]
f_b(z) &\simeq& -1.836 + 2.608\, z + 0.8271\, z^2 - 2.441\, z \ln z,\nnb\\[1mm]
f_c(z) &\simeq&  9.099 + 13.20\, z -  19.68\, z^2 + 25.71\, z \ln z,\label{4funct}
\eea
of $z = m_c^2/m_b^2$ and $\delta = 1 - 2E_0/m_b$ are going to be useful.
Explicit formulae for $r_2^{(1)}(z)$ and ${\rm Re}\,r_2^{(2)}(z)$ can be found
in Eq.~(3.1) of Ref.~\cite{Buras:2002tp} and Eq.~(26) of
Ref.~\cite{Bieri:2003ue}, respectively. For $h^{(2)}_{27}(z,\delta)$, we shall use a
numerical fit from Eq.~(13) of Ref.~\cite{Misiak:2010tk}. An analytical
expression for $\phi^{(1)}_{27}(z,\delta)$~ for~ $4z < 1-\delta$~ (which is
the phenomenologically relevant region) reads
\bea
\phi^{(1)}_{27}(z,\delta) &=& -\f{2}{27} \delta (3-3\delta+\delta^2) + 
\f43 z\, \delta\, s_\delta\, L_\delta + \f{12-8\pi^2}{9} z^2 \delta +
\f43 z (1-2z) ( s_0 L_0 - s_\delta L_\delta )\nnb\\[2mm]
&+& \f{2\pi^2-7}{9} z \delta (2-\delta) 
- \f89 z (6z^2-4z+1)( L_0^2 - L_\delta^2 ) 
-\f89 z \delta (2-\delta-4z) L_\delta^2, \label{phi27}
\eea
with~ $s_\delta = \sqrt{(1-\delta)(1-\delta-4z)}$,~ $s_0 = \sqrt{1-4z}$,~
$L_\delta = \ln\f{\sqrt{1-\delta}+\sqrt{1-\delta-4z}}{2\sqrt{z}}$~
and~ $L_0 = \ln\f{1+\sqrt{1-4z}}{2\sqrt{z}}$.

In the $\delta=1$ case, $\phi^{(1)}_{27}$ and $h^{(2)}_{27}$ 
for $z < \f14$ are given by
\mathindent0cm
\bea
\phi^{(1)}_{27}(z,1) &=& -\f{2}{27} + \f{12-8\pi^2}{9} z^2 +
\f43 z (1-2z) s_0 L_0 + \f{2\pi^2-7}{9} z 
- \f89 z (6z^2-4z+1) L_0^2 + \f43 \pi^2 z^3,\nnb\\[2mm]
h^{(2)}_{27}(z,1) &\simeq& \f{41}{27} - \f29 \pi^2 - 2.24\, z^{1/2} - 7.04\, z +
       23.72\, z^{3/2} + ( -9.86\, z + 31.28\, z^2) \ln z. 
\eea
\mathindent1cm

The functions $f_b(z)$ and $f_c(z)$ in Eq.~(\ref{4funct}) come from Eqs.~(3.3)
and (3.4) of Ref.~\cite{Boughezal:2007ny}, respectively. These numerical fits
(in the range $z \in [ 0.017, 0.155 ]$) describe contributions from three-loop
$b \to s\gamma$ amplitudes with massive $b$-quark and $c$-quark loops on the
gluon lines.

The ratio $z = m_c^2/m_b^2$ is defined in terms of the $\overline{\rm
MS}$-renormalized charm quark mass at an arbitrary scale $\mu_c$. In practice,
we shall use $\mu_c=2.0\,$GeV as a central value.  As far as the
renormalization scheme for $m_b$ is concerned, we assume the following relation
to the on-shell scheme 
\be \f{m_{b,\rm pole}}{m_b} = 1 + \alt x_m + {\mathcal O}(\alt^2).  
\ee 
In the 1S and kinetic schemes, one finds $x_m = \fm{8}{9} \pi \alpha_\Upsilon$ 
and $x_m = \f{64 \mu_{\rm kin}}{9 m_b} \left( 1 + \f{3 \mu_{\rm kin}}{8 m_b} \right)$, 
respectively. In our numerical analysis, the kinetic scheme is going to be used.

Complete expressions for the NNLO quantities $K^{(2)}_{17}$ and $K^{(2)}_{27}$ 
can now be written as follows
\bea
K^{(2)}_{17}(z,\delta) &=& -\f{1}{6} K^{(2)}_{27}(z,\delta) + A_1 + F_1(z,\delta) +
\left( \f{94}{81} - \f{3}{2} K^{(1)}_{27} - \f{3}{4} K^{(1)}_{78} \right) L_b - \f{34}{27} L_b^2,\nnb\\
K^{(2)}_{27}(z,\delta) &=& A_2 + F_2(z,\delta) 
-\f32 \beta_0^{n_l=3} f_q(z,\delta) + f_b(z) + f_c(z) + \f{4}{3} \phi^{(1)}_{27}(z,\delta) \ln z \nnb\\
&+& \left[ (8L_c - 2x_m)\, z \f{d}{dz} \;+\; 
    (1-\delta) x_m \f{d}{d\delta} \right] f_{\scs NLO}(z,\delta) + \f{416}{81} x_m \nnb\\
&+& \left( \f{10}{3} K^{(1)}_{27} - \f{2}{3} K^{(1)}_{47} 
     - \f{208}{81} K^{(1)}_{77} - \f{35}{27} K^{(1)}_{78} - \f{254}{81} \right) L_b - \f{5948}{729} L_b^2,
\label{k27nnlo}
\eea
where~ $\beta_0^{n_l=3} = 9$,
$L_b = \ln(\mu_b^2/m_b^2)$~ and~ $L_c = \ln(\mu_c^2/m_c^2)$,~ while 
the relevant $K_{ij}^{(1)}$ are collected in Appendix~C.

The expressions~ $A_i + F_i(z,\delta)$~ contain all the contributions that are
not yet known for the measured value of $m_c$. They correspond to those parts of
the considered interference terms that are obtained by:
{\it (i)} setting $\mu_b = m_b$, $\mu_c = m_c$ and $x_m = 0$,
{\it (ii)} removing the BLM-extended contributions from quark loops on the gluon lines and from
           $b \to s q \bar q \gamma$ decays ($q=u,d,s$), except for those given 
           in Fig.~\ref{fig:ab}b.

We define the constants $A_i$ by requiring that $F_i(0,1)=0$. Then we evaluate
$A_i$ from Eq.~(\ref{renormalized}) by setting there $\mu = m_b$, $n_b=0$ and
$\kappa n_l = 3$. Next, a replacement $n_l \to n_l + \f32 \beta_0^{n_l} =
\f{33}{2}$ is done in the remaining $n_l$-terms.  Finally, Eq.~(\ref{G2K})
is used to find
\be
A_1 \simeq 22.605, \hspace{2cm} A_2 \simeq 75.603.
\ee
These two numbers are the only outcome of our calculation in Section~\ref{sec:zero.mc}
that is going to be used in the phenomenological analysis below.

Apart from the condition $F_i(0,1)=0$, everything that is known at the moment
about the functions $F_i(z,\delta)$ are their large-$z$ asymptotic forms. They
can be derived from the results of Ref.~\cite{Misiak:2010sk}.\footnote{
We supplement them now with the previously omitted large-$m_c$ contributions
from the diagrams in Fig.~1 in Ref.~\cite{Misiak:2010sk}
or, equivalently, Fig.~\ref{fig:ab}b in the present paper.
The effect of such a modification is numerically very small.}
Explicitly, we find
\bea
F_1(z,\delta) &=& \f{70}{27} \ln^2 z + \left( \f{119}{27} - \f{2}{9}\pi^2 
    + \f{3}{2} \phi^{(1)}_{78}(\delta) \right) \ln z - \f{493}{2916} - \f{5}{54} \pi^2 
    + \f{232}{27} \zeta_3 + \f{5}{8} \phi^{(1)}_{78}(\delta)\nnb\\[1mm]
&-& A_1 + {\mathcal O}\left(\f{1}{z}\right),\nnb\\[2mm]
    F_2(z,\delta) &=&  -\f{4736}{729} \ln^2 z + \left\{ 
-\f{165385}{2187} + \f{1186}{729} \pi^2 - \f{2 \pi}{9 \sqrt{3}} + \f23 Y_1
     + \f{4}{3} \phi^{(1)}_{47}(\delta) + \f{832}{81} \phi^{(1)}_{77}(\delta)\right.\nnb\\[1mm]
&+& \left. \f{70}{27} \phi^{(1)}_{78}(\delta)  \right\} \left(\ln z + 1\right) -\f{956435}{19683} 
    - \f{2662}{2187} \pi^2 + \f{20060}{243} \zeta_3 - \f{1624}{243} \phi^{(1)}_{77}(\delta)\nnb\\[1mm]  
&-& \f{293}{162} \phi^{(1)}_{78}(\delta) - A_2 + {\mathcal O}\left(\f{1}{z}\right).\label{asymp}
\eea
The constant $Y_1$ and the necessary $\phi^{(1)}_{ij}$ functions are given
in Appendices~B and C, respectively.

Let $\Delta{\mathcal B}_{s\gamma}$ denote the contribution from $F_{1,2}(z,\delta)$ to
${\mathcal B}_{s\gamma}$. Then the relative effect is given by
\be
\f{\Delta{\mathcal B_{s\gamma}}}{\mathcal B_{s\gamma}} ~\simeq~ U(z,\delta) ~\equiv~
\f{\al^2(\mu_b)}{8\pi^2}\;\; \f{ C_1^{(0)}(\mu_b) F_1(z,\delta) + \left( C_2^{(0)}(\mu_b)
-\f{1}{6}C_1^{(0)}(\mu_b)\right) F_2(z,\delta)}{C_7^{(0)\rm eff}(\mu_b)}.
\ee
For $\mu_b=2.0\,$GeV, we have $\alpha_s(\mu_b) \simeq 0.293$,
$C_1^{(0)}(\mu_b) \simeq -0.902$, $C_2^{(0)}(\mu_b) \simeq 1.073$,
and $C_7^{(0)\rm eff}(\mu_b) \simeq -0.385$.

We shall estimate the contribution to ${\mathcal B}_{s\gamma}$ that comes from the
unknown $U(z,\delta)$ by considering an interpolation model where
$U(z,1)$ is given by the following linear combination
\be \label{uinterp}
U_{\rm interp}(z,1) ~=~ x_1 \,+\, x_2\, f_q(z,1) \,+\, \left( x_3 \,+\, x_4\, z \f{d}{dz} \right) f_{\scs NLO}(z,1).
\ee
The numbers $x_i$ are fixed by the condition $U(0,1)=0$ as well as by the
large-$z$ behaviour of $U(z,1)$ that follows from Eq.~(\ref{asymp}).  This
determines $x_i$ in a unique manner, namely $x_i \simeq 
(-0.0502,~0.0328,~0.0373,~0.0309)_i$. In Fig.~\ref{fig:interp}, the
function $U_{\rm interp}(z,1)$ is plotted with a solid line, while the dashed
line shows $U_{\rm asymp}(z,1)$, i.e.\ asymptotic large-$z$ behaviour of the
true $U(z,1)$. Note that $\sqrt{z}=m_c/m_b$ rather than $z$ is used on
the horizontal axis.  The vertical line corresponds to the measured value of
this mass ratio. The plot involves some extra approximation in the region between
$\sqrt{z} \simeq 0.4$ and $\sqrt{z} \simeq 0.8$ where we need to interpolate
between the known small-$z$ and large-$z$ expansions of ${\rm
Re\,}r_2^{(2)}(z)$ (see Fig.~1 of Ref.~\cite{Misiak:2006ab}).

In Refs.~\cite{Misiak:2006ab,Misiak:2006zs} the uncertainty in ${\mathcal
B}_{s\gamma}$ due to unknown $m_c$-dependence of the NNLO corrections
has been estimated at the $\pm 3\%$ level. The size of the interpolated
contribution in Fig.~\ref{fig:interp} implies that no reduction of this
uncertainty is possible at the moment. One might wonder whether the
uncertainty should not be enlarged.  Our choice here is to leave it
unchanged, for the following reasons:
\begin{figure}[t]
\begin{center}
\includegraphics[width=80mm,angle=0]{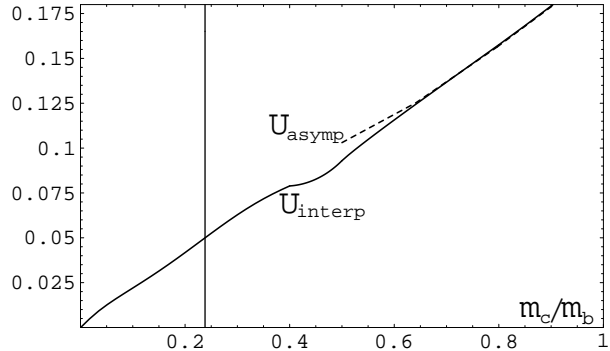}
\caption{\sf The interpolating function defined in Eq.~(\ref{uinterp}) (solid line) 
and asymptotic behaviour of the true function $U(z,1)$ for $m_c \gg m_b/2$ (dashed line). 
The vertical line corresponds to the measured value of $m_c/m_b$.
\label{fig:interp}}
\end{center}
\end{figure}
\begin{itemize}
\item[(i)] Our choice of functions for the linear combination in
Eq.~(\ref{uinterp}) is dictated by the fact, that these very functions
determine the dependence on $z$ of the known parts of $K^{(2)}_{17}$
and $K^{(2)}_{27}$. The known parts are either those related to
renormalization of the Wilson coefficients and quark masses (in the terms
proportional to $L_b$ and $L_c$) or the renormalization of $\alpha_s$ (the
function $f_q$ parametrizes the considered correction in the BLM
approximation).  It often happens in perturbation theory that higher-order
corrections are dominated by renormalization effects. If this is the case
here, the true $U(z,1)$ should have a similar shape to $U_{\rm interp}(z,1)$.
\item[(ii)] The growth of $U_{\rm interp}(z,1)$ for $m_c >
m_b/2$ is perfectly understandable. In this region, logarithms of $z$ from
Eq.~(\ref{asymp}) combine with $L_b$ from Eq.~(\ref{k27nnlo}), and the
asymptotic large-$m_c$ behaviour of $K^{(2)}_{(1,2)7}$ is determined by
$\ln(\mu_b/m_c)$ and $\ln(\mu_c/m_c)$ only (see Eqs.~(5.12) and (5.14)
of Ref.~\cite{Misiak:2010sk}). Thus, the growth of the correction for large
$z$ can be compensated by an appropriate choice of the renormalization scales,
which means (not surprisingly) that the dangerous large logarithms can get
resummed using renormalization group evolution of the Wilson coefficients,
masses and $\alpha_s$.
\item[(iii)] Our $\pm 3\%$ uncertainty is going to be combined in quadrature
with the other ones, which means that it should be treated as a ``theoretical
$1\sigma$ error''. To gain higher confidence levels, it would need to be enlarged.
\item[(iv)] In the considered interference terms $K_{17}$ and
$K_{27}$, the dependence on $\delta$ is very weak in the whole range
$\delta \in [0,1]$, both at the NLO and in the BLM approximation for the NNLO
corrections.  Specifically, changing $\delta$ from 1 ($E_0 = 0$) to 0.295
($E_0 = 1.6\,$GeV) results in modifications of $f_{NLO}$ by $+0.2\%$ and $f_q$
by $+1.0\%$, respectively, for the measured value of $m_c$. The corresponding
changes at $m_c=0$ amount to $-0.7\%$ and $-2.4\%$ only. Thus, our estimates
made for $\delta=1$ are likely to be valid for arbitrary $\delta$.
\end{itemize}

In the phenomenological analysis below, we shall take $K^{(2)}_{17}$ and
$K^{(2)}_{27}$ as they stand in Eq.~(\ref{k27nnlo}), replace the unknown
$F_i(z,\delta)$ by $F_i^{\rm interp}(z,1)$ interpolated analogously to
Eq.~(\ref{uinterp})
\bea 
F_1^{\rm interp}(z,1) &=& -23.75 \,+\, \f{35}{12}\, f_q(z,1) \,+\, 
\left( \f{2129}{936} - \f{9}{52} \pi^2 \,-\, 0.84\, z \f{d}{dz} \right) f_{\scs NLO}(z,1),\nnb\\[2mm]
F_2^{\rm interp}(z,1) &=& -3.01 \,-\, \f{592}{81}\, f_q(z,1) \,+\, 
\left( -10.34 \,-\, 9.55\, z \f{d}{dz} \right) f_{\scs NLO}(z,1),
\label{finterp}
\eea
and include a $\pm 3\%$ uncertainty in the branching ratio due to such an
approximation. 

\newsection{Evaluation of ${\mathcal B}_{s\gamma}$ in the SM \label{sec:pheno}}

In the present section, we include all the other corrections to ${\mathcal
B}_{s\gamma}$ that have been evaluated after the analysis in
Refs.~\cite{Misiak:2006ab,Misiak:2006zs}. Next, we update the SM
prediction. To provide information on sizes of the subsequent corrections, the
description is split into steps, and the corresponding modifications in the
branching ratio central value are summarized in Tab.~\ref{tab:brshifts}.  The
steps are as follows:
\begin{itemize}
\item[1.] We begin with performing the calculation precisely as it was
          described in Ref.~\cite{Misiak:2006ab} but only shifting from
          ${\mathcal B}(\bar B \to X_s \gamma)$ to ${\mathcal B}_{s\gamma}$,
          which amounts to CP-averaging the perturbative decay widths. No
          directly CP-violating non-perturbative corrections to ${\mathcal
          B}(\bar B \to X_s \gamma)$ were considered in
          Ref.~\cite{Misiak:2006ab}. It was not equivalent to neglecting them
          but rather to assuming that they have vanishing central values. A
          dedicated analysis in Ref.~\cite{Benzke:2010tq} leads to an estimate
          of~ $0.4 \pm 1.7\%$~ for such effects.
%
%
\begin{table}[t]
\begin{center}
\begin{tabular}{|cccccccccc|c|}\hline
1 & 2 & 3 & 4 & 5 & 6 & 7 & 8 & 9 & 10 & total \\\hline
  $-0.6\%\!$ 
& $+1.0\%\!$ 
& $-0.2\%\!$ 
& $+2.0\%\!$ 
& $+1.0\%\!$ 
& $+1.6\%\!$ 
& $+2.1\%\!$ 
& $-0.5\%\!$ 
& $+0.2\%\!$ 
& $-0.4\%\!$ 
& $+6.4\%\!$ 
\\\hline
\end{tabular}
\end{center}
\ \\[-1cm]
\caption{\sf Shifts in the central value of ${\mathcal B}_{s\gamma}$
for $E_0 = 1.6\,$GeV at each step (see the text).\label{tab:brshifts}}
\end{table}
\item[2.] The input parameters are updated as outlined in
                 Appendix~D. In particular, we use results of the very recent
                 kinetic-scheme fit to the semileptonic $B$ decay
                 data~\cite{Alberti:2014yda}.
\item[3.] Central values of the renormalization scales
                 $(\mu_c,\mu_b)$ are shifted from $(1.5,2.5)\,$GeV to
                 $(2,2)\,$GeV. Both scales are then varied in the ranges
                 $[1.25,~5]\,$GeV to estimate the higher-order uncertainty.
                 In the resulting range of ${\mathcal B}_{s\gamma}$, the
                 value corresponding to the $(2,2)\,$GeV renormalization scales
                 is more centrally located than the $(1.5,2.5)\,$GeV one,
                 after performing all the updates 1-10 here. It is the main 
                 reason for shifting the default scales. The $(2,2)\,$GeV choice is 
                 also simpler (both scales are equal), and $\mu_c$ is exactly
                 as in the fit from which we take $m_c(\mu_c)$ (Appendix~D).
                 As far as $\mu_b$ is concerned, it should be of the same order 
                 as the energy transferred to the partonic system after the $b$-quark decay.
                 For the leading $b\to s\gamma$ contribution from the photonic dipole
                 operator $P_7$, this energy equals to $\f12 m_b$ which gives $2.3\,$GeV
                 when one substitutes $m_b = m_{b,kin}$ from Appendix~D.\footnote{ 
The measured photon spectra are also peaked at around $2.3\,$GeV, which confirms 
    the leading role of the two-body partonic mode.}
                Rounding 2.3 to either 2.5 or 2.0 for the default value is
                equally fine, given that the observed $\mu_b$-dependence of
                ${\mathcal B}_{s\gamma}$ is weak (see Fig.~\ref{fig:mudep}),
                and our range for $\mu_b$ is $[1.25,~5]\,$GeV.
\item[4.] In the interpolation of $P_2^{(2)\rm rem}$ (see
                 Ref.~\cite{Misiak:2006ab} for its definition), we shift to
                 the so-called case~(c) where the interpolated quantity at
                 $m_c=0$ was given by the $(Q_7,Q_7)$ interference alone.
\item[5.] The $m_c=0$ boundary for $P_2^{(2)\rm rem}$ is updated to
                 include all the relevant interferences, especially the ones
                 evaluated in Section~\ref{sec:zero.mc}.  The thick solid
                 (red) line in Fig.~\ref{fig:p22rem} shows the new
                 $P_2^{(2)\rm rem}$ in such a case, while the remaining lines
                 are as in Fig.~2 of Ref.~\cite{Misiak:2006ab} (somewhat
                 shifted due to the parameter and scale modifications only).  
\begin{figure}[t]
\begin{center}
\includegraphics[width=81mm,angle=0]{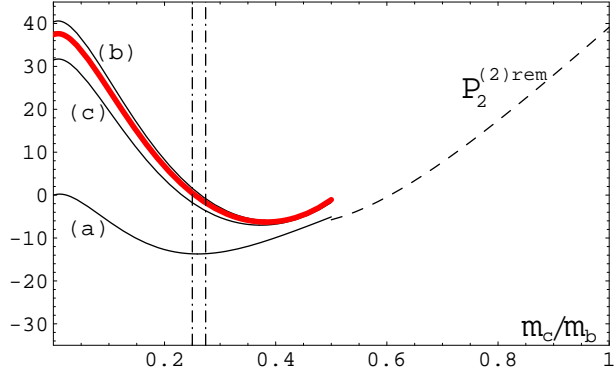}
\caption{\sf Interpolation of $P_2^{(2)\rm rem}$ in $m_c$ as in Fig.~2 of
Ref.~\cite{Misiak:2006ab} but with updated input parameters and with
renormalization scales shifted to $(\mu_c,\mu_b) = (2,2)\,$GeV. In
addition, the thick solid (red) line shows the case with the presently known
boundary condition at $m_c=0$ imposed.\label{fig:p22rem}}
\end{center}
\end{figure}
\item[6.] At this point, we abandon the approach with
                 $m_c$-interpolation applied to the {\em whole} non-BLM
                 correction $P_2^{(2)\rm rem}$.  As before, the penguin
                 operators $Q_{3,\ldots,6}$ and the CKM-suppressed ones
                 $Q^u_{1,2}$ are neglected at the NNLO level. The
                 corrections $K^{(2)}_{17}$ and $K^{(2)}_{27}$ are treated as
                 summarized at the end of the previous section. For
                 $K^{(2)}_{78}$, the complete results from
                 Refs.~\cite{Ewerth:2008nv,Asatrian:2010rq} are included.
                 $K^{(2)}_{77}$ is made complete by taking into account its
                 exact $m_c$-dependence~\cite{Asatrian:2006rq,Pak:2008qt}, in
                 addition to the previously included terms.  For the NNLO
                 interferences among $Q_1$, $Q_2$ and $Q_8$, only the two-body
                 final state contributions are present at this step. They are
                 infrared-finite by themselves, and given by products of the
                 well-known NLO amplitudes $r_i^{(1)}$ (see Eq.~(3.1) of
                 Ref.~\cite{Buras:2002tp}) whose imaginary parts matter here,
                 too.
\item[7.] Three- and four-body final state contributions to the NNLO
                  interferences among $Q_1$, $Q_2$ and $Q_8$ are included in
                  the BLM approximation, using the results of
                  Refs.~\cite{Ligeti:1999ea,Ferroglia:2010xe,Misiak:2010tk}.
                  Non-BLM corrections to these interferences remain neglected.
%
%
                  The corresponding uncertainty is going to be absorbed below into 
                  the overall $\pm 3\%$ perturbative one.
\item[8.] Four-loop $Q_{1,\ldots,6} \to Q_8$ anomalous dimensions from
                 Ref.~\cite{Czakon:2006ss} are included in the renormalization
                 group equations.
\item[9.] The LO and NLO contributions from four body final states are
               included~\cite{Kaminski:2012eb,Huber:2014nna}.  They are not
               yet formally complete, but the only neglected terms are the NLO
               ones that undergo double (quadratic) suppression either by the
               small Wilson coefficients $C_{3,\ldots,6}$ or by the small CKM
               element ratio $\left|V_{us}^* V_{ub}\right|/\left|V_{ts}^*
               V_{tb}\right|$. The uncertainty that results from neglecting
               such terms is below a permille in ${\mathcal
               B}_{s\gamma}$. As far as the CKM-suppressed two-body and
               three-body contributions are concerned, the two-body NLO one
               has already been taken into account in
               Ref.~\cite{Misiak:2006ab}.  The remaining NLO and NNLO ones
               (also those with double CKM suppression) are included at the
               present step. Their contribution to ${\mathcal B}_{s\gamma}$ is
               below a permille. However, the branching ratio ${\mathcal
               B}_{d\gamma}$~\cite{Misiak:2015xxx} receives around 2\%
               enhancement from them.
\item[10.] We update our treatment of non-perturbative
          corrections. The ${\mathcal O}\left(\al\Lambda^2/m_b^2\right)$ 
          correction to the $(Q_7,Q_7)$ interference from
          Ref.~\cite{Ewerth:2009yr} replaces the previous approximate
          expression from Ref.~\cite{Neubert:2004dd}.  Moreover, we
          include a similar correction~\cite{Alberti:2013kxa,Gambino:2014priv} 
          to the charmless semileptonic rate that is used for normalization
          in $[P(E_0) + N(E_0)]$ (see Eqs.~(\ref{phase}) and (\ref{brB})
          in Appendix~D). In consequence, the previous (tiny) effect in
          $N(E_0)$ gets reduced by a factor of around 4.
           Finally, our treatment of non-perturbative effects in interferences
           other than $(Q_7,Q_7)$ gets modified according to
           Ref.~\cite{Benzke:2010js}. A vanishing contribution to the
           branching ratio central value from such corrections is assumed,
           except for the leading ${\mathcal O}\left(\lambda_2/m_c^2\right)$ 
           one~\cite{Buchalla:1997ky} where $m_c$ is fixed to $1.131\,$GeV. At
           the same time, a $\;\pm 5\%$ non-perturbative uncertainty in the
           branching ratio is assumed, as obtained in Sec.~7.4 of
           Ref.~\cite{Benzke:2010js} by adding the relevant three
           uncertainties in a linear manner.\footnote{
If their ranges were treated as $1\sigma$ ones and combined in quadrature, the
uncertainty would go down to $3.3\%$.}
\end{itemize}
\begin{figure}[t]
\begin{center}
\includegraphics[width=8cm,angle=0]{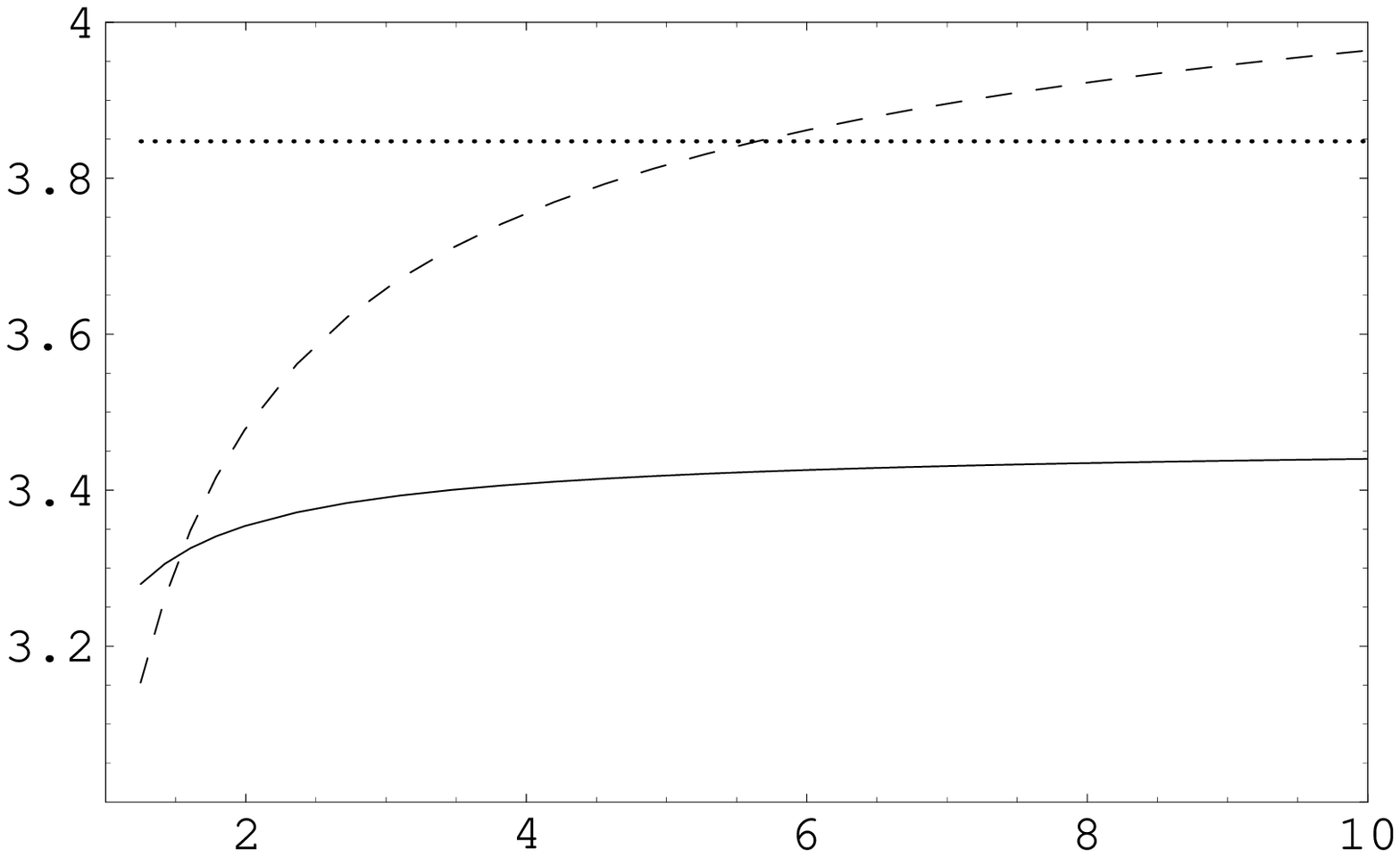}
\hspace{5mm}
\includegraphics[width=8cm,angle=0]{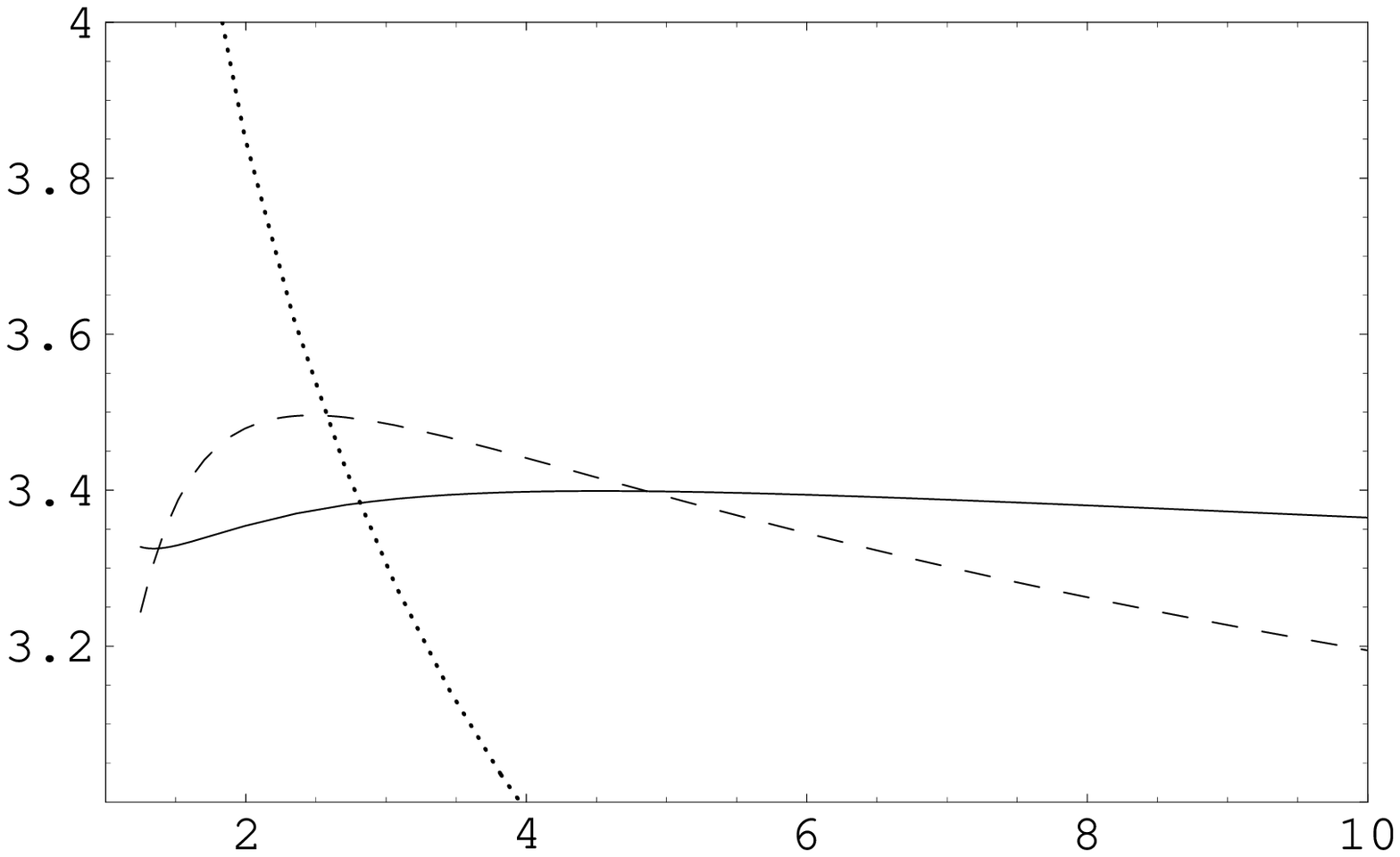}\\
\includegraphics[width=8cm,angle=0]{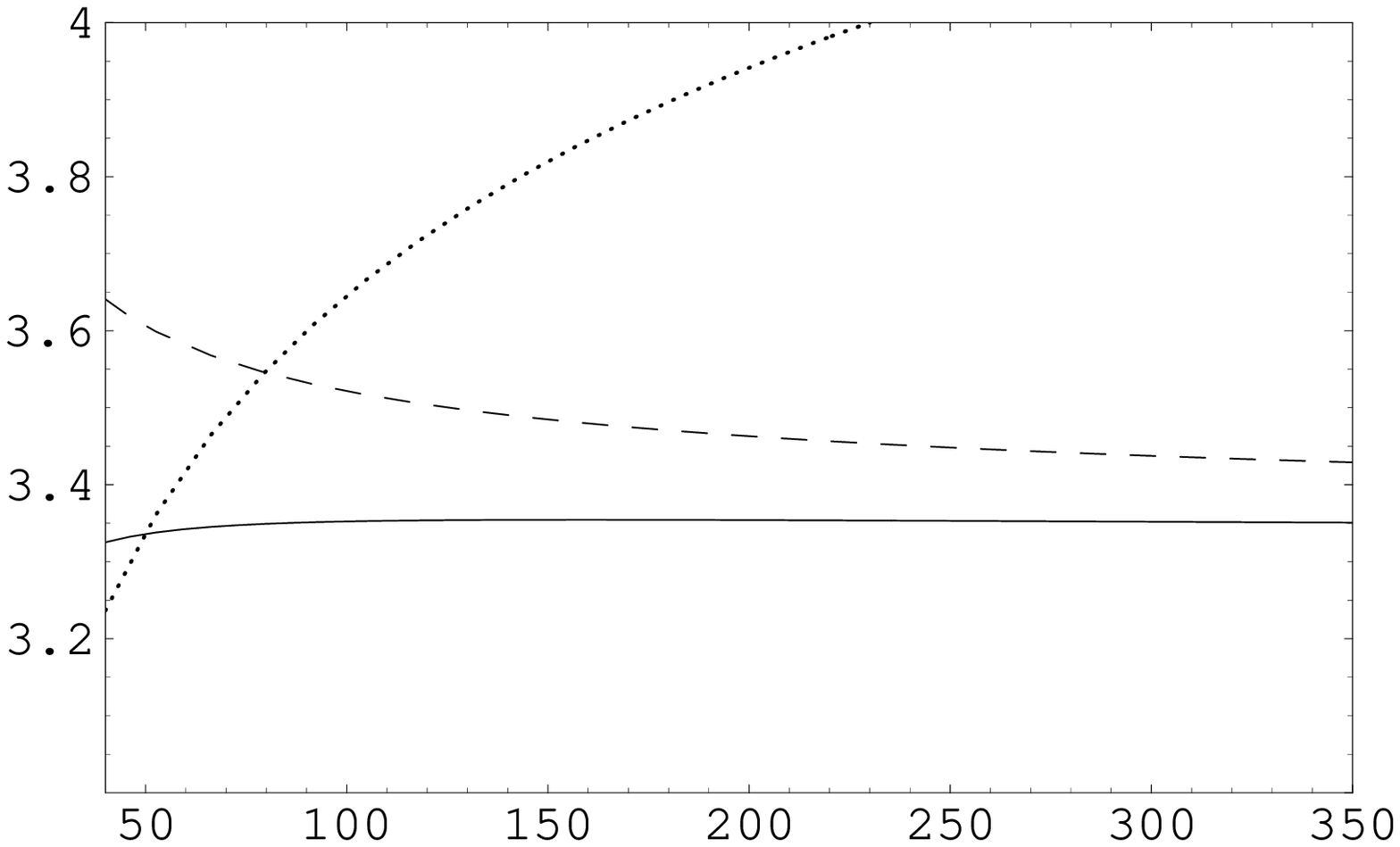}
\end{center}
\vspace*{-67mm} 
\hspace*{39mm} $\mu_c$ \hspace{81mm} $\mu_b$\\[46mm]
\hspace*{86mm} $\mu_0$\\ 
\begin{center}
\caption{\sf Renormalization scale dependence of ${\cal B}_{s\gamma}$
  in units $10^{-4}$ at the LO (dotted lines), NLO (dashed lines) and NNLO (solid
  lines). The upper-left, upper-right and lower plots describe the dependence
  on $\mu_c$, $\mu_b$ and $\mu_0$ [GeV], respectively. When one of the scales
  is varied, the remaining ones are set to their default values.\label{fig:mudep}}
\end{center}
\end{figure}

Our final result reads
\be \label{finalBR}
{\mathcal B}^{\rm SM}_{s\gamma} = (3.36 \pm 0.23) \times 10^{-4}
\ee
for~ $E_0 = 1.6\,$GeV, where four types of uncertainties have been combined in
quadrature:\linebreak $\pm 5\%$ non-perturbative (step 10 above), 
$\pm 3\%$ from our interpolation of $F_{1,2}(z,\delta)$
(Section~\ref{sec:mcdep}), $\pm 2.0\%$ parametric (Appendix~D), as
well as $\pm 3\%$ from higher-order perturbative effects.

The latter uncertainty is assumed to account for approximations made at the
NLO and NNLO levels, too. In the NLO case, it refers to the doubly
suppressed terms mentioned in step 9 above. In the NNLO case, it refers to
neglecting the penguin operators at this level, and using the BLM
approximation in step 7 above. If we relied just on the
renormalization-scale dependence in Fig.~\ref{fig:mudep} (with $1.25\,{\rm
GeV} < \mu_c,\mu_b < 5\,$GeV), we could reduce this uncertainty to around 
$\pm 2.4\%$. However, apart from the scale-dependence, one needs to
study how the perturbation series behaves, which is hard to judge before
learning the actual contributions from $F_{1,2}(z,\delta)$. Thus, we leave the
higher-order uncertainty unchanged with respect to
Refs.~\cite{Misiak:2006ab,Misiak:2006zs}. Our treatment of the electroweak
corrections~\cite{Gambino:2001au} remains unchanged, too.

The central value in Eq.~(\ref{finalBR}) is about 6.4\% higher than
the previous estimate of $3.15\times 10^{-4}$ 
%
%
in Refs.~\cite{Misiak:2006ab,Misiak:2006zs}. Around half of this effect
comes from improving the $m_c$-interpolation. As seen in
Fig.~\ref{fig:p22rem}, the currently known $m_c=0$ boundary for the thick
line is close to the edge of the previously assumed range between
the curves (a) and (b). It is consistent with the fact that the corrections in steps 
4 and 5 sum up to 3\% being the previous ``$1\sigma$'' interpolation uncertainty.
The $m_c=0$ boundary has been the main worry in the past because 
estimating the range for its location was based on quite arbitrary
assumptions. It is precisely the reason why no update of the SM prediction
seemed to make sense until now, given moderate sizes of the other new
corrections.

\newsection{Conclusions \label{sec:concl}}

We evaluated ${\mathcal O}(\alpha_s^2)$ contributions to the perturbative
$\Gamma(b \to X_s \gamma)$ decay rate that originate from the $(Q_7,Q_{1,2})$
interference for $m_c = E_0 = 0$. The calculation involved 163 four-loop
massive on-shell propagator master integrals with unitarity cuts. Our updated
prediction for the CP- and isospin-averaged branching ratio in the SM reads~
${\mathcal B}^{\rm SM}_{s\gamma} = (3.36 \pm 0.23) \times 10^{-4}$. It
includes all the perturbative and non-perturbative contributions that have
been calculated to date. It agrees very well with the current 
experimental world average ${\mathcal B}^{\rm exp}_{s\gamma} = (3.43
\pm 0.21 \pm 0.07) \times 10^{-4}$. An extension of our analysis to the case
of ${\mathcal B}_{d\gamma}$ and an update of bounds on the Two Higgs Doublet
Model is going to be presented in a parallel article~\cite{Misiak:2015xxx}.

\section*{Acknowledgments}

We would like to thank Ulrich Nierste for helpful discussions, and
Paolo Gambino for extensive correspondence concerning non-perturbative
corrections and input parameters, as well as for providing us with the
semileptonic fit results in an unpublished option (see Appendix~D). We are
grateful to Micha{\l} Poradzi\'nski and Abdur Rehman for performing several
cross-checks of the three- and four-body contributions.  The work of M.C.,
P.F.\ and M.S.\ has been supported by the Deutsche Forschungsgemeinschaft in
the Sonderforschungsbereich Transregio~9 ``Computational Particle
Physics''. T.H.\ acknowledges support from the Deutsche Forschungsgemeinschaft
within research unit FOR 1873 (QFET). M.M.\ acknowledges partial support by
the National Science Centre (Poland) research project, decision no
DEC-2014/13/B/ST2/03969.

\newappendix{Appendix~A:~~ Massless master integrals}
\def\theequation{A.\arabic{equation}}
\begin{figure}[t]
\begin{center}
\begin{tabular}{cccc}
\includegraphics[width=39mm,angle=0]{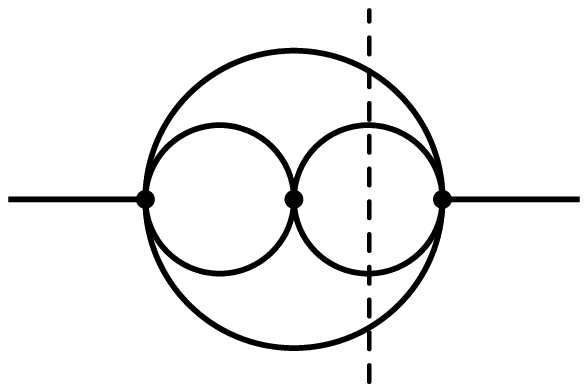} &\includegraphics[width=39mm,angle=0]{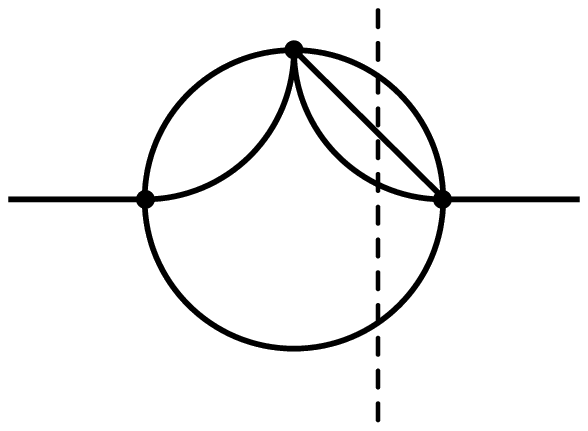}&
\includegraphics[width=39mm,angle=0]{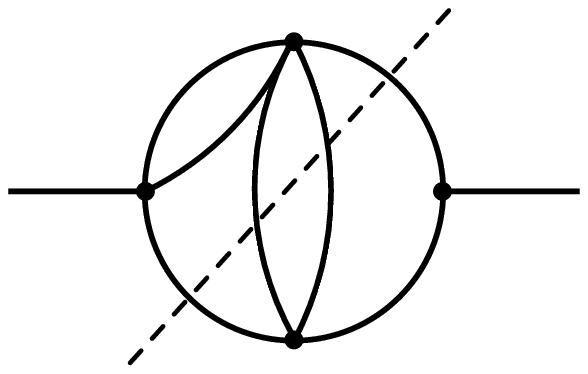} &\includegraphics[width=39mm,angle=0]{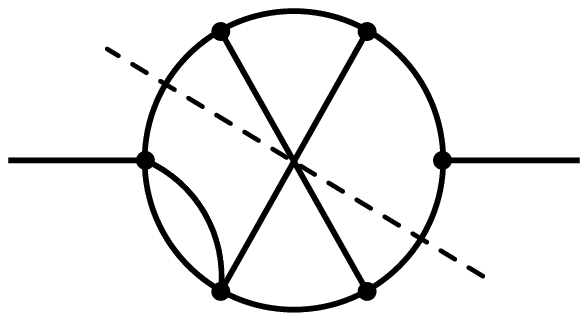}\\
4L4C1&4L4C2&4L4C3&4L4C4\\
\includegraphics[width=39mm,angle=0]{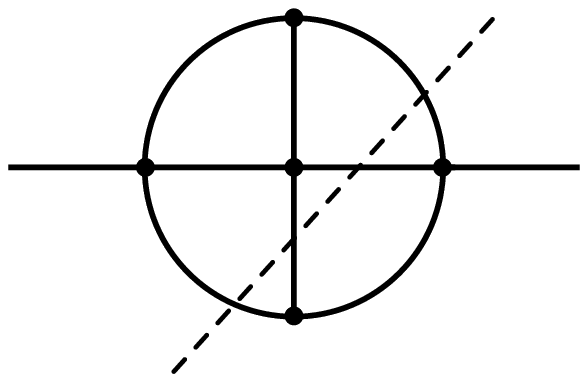} &\includegraphics[width=39mm,angle=0]{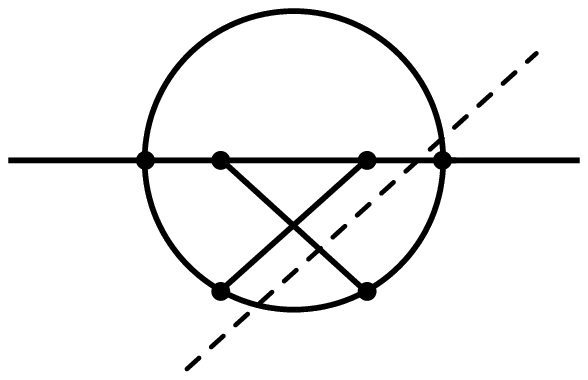}&
\includegraphics[width=39mm,angle=0]{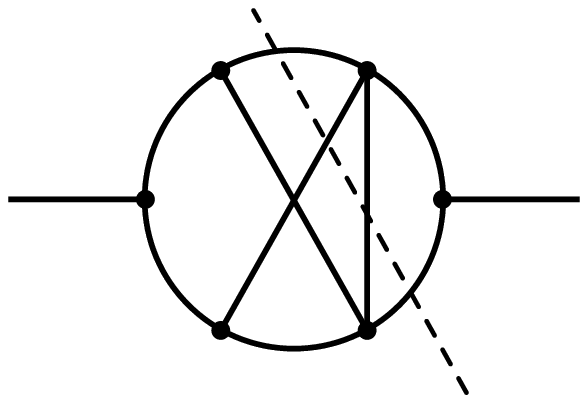} &\includegraphics[width=39mm,angle=0]{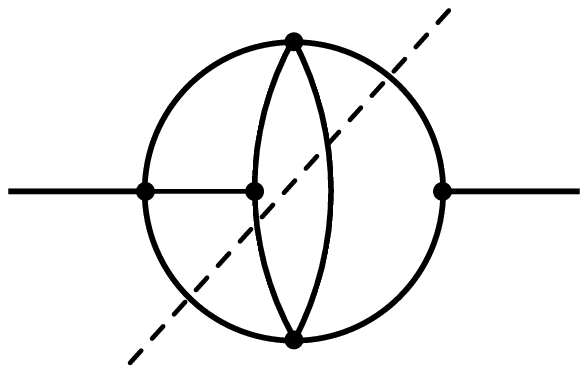}\\
4L4C5&4L4C6&4L4C7&4L4C8
\end{tabular}
\end{center}
\caption{\sf The massless four-particle-cut diagrams calculated in the course of this work.  \label{fig:4PCutMasters}}
\end{figure}

In the course of this work, it has been necessary to compute a number of
massless scalar integrals with various unitarity cuts. All of them are
depicted in Fig.~\ref{fig:4PCutMasters} and Tab.~\ref{tab:masters}.  They
occur after applying the large mass expansion for $p_b^2 \ll m_b^2$, as well
as in the decay rate calculation itself. Apart from the four-loop diagrams
with four-particle cuts, and the four-loop diagrams 4L3C1, 4L3C2 and 4L3C3
with three-particle cuts, values of all our master integrals can either be
found in the 
literature~\cite{Huber:2010fz,GehrmannDeRidder:2003bm,Baikov:2009bg,Lee:2010cga,Gehrmann:2010ue}
or obtained using standard techniques described, for instance, in
Ref.~\cite{smirnovbook}. Let us note that the results for all the massless
propagator four-loop master integrals in
Refs.~\cite{Baikov:2010hf,Smirnov:2010hd} are not sufficient here because they
correspond to sums over all the possible cuts, while certain cuts need to be
discarded in our case.
\begin{table}[t]
\begin{center}
\begin{tabular}{|cc|ccc|}
\hline
\multicolumn{2}{|c|}{2PCuts}&\multicolumn{3}{|c|}{3PCuts}\\
\hline
\multicolumn{2}{|c|}{\includegraphics[width=21mm,angle=0]{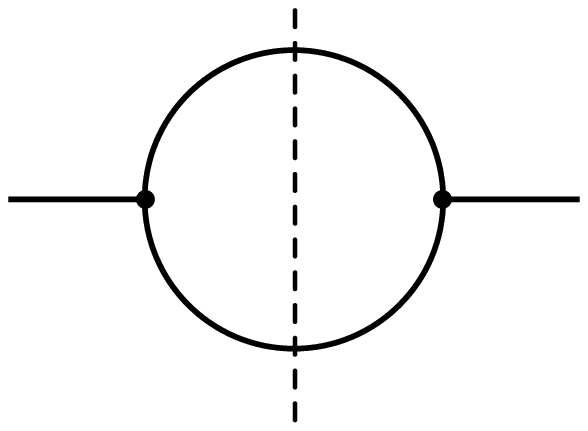}} &&&\\
\multicolumn{2}{|c|}{1L2C1} &&&\\
\hline
\multicolumn{2}{|c|}{\includegraphics[width=21mm,angle=0]{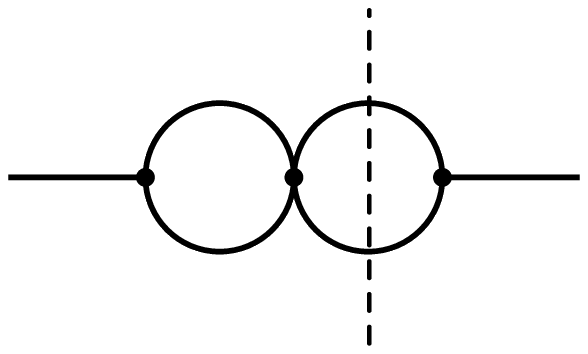}}&
\multicolumn{3}{|c|}{\includegraphics[width=21mm,angle=0]{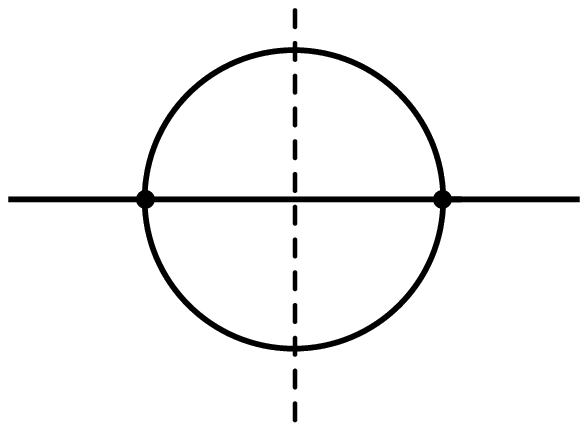}}\\
\multicolumn{2}{|c|}{2L2C1}&\multicolumn{3}{|c|}{2L3C1}\\
\hline
\multicolumn{2}{|c|}{\includegraphics[width=21mm,angle=0]{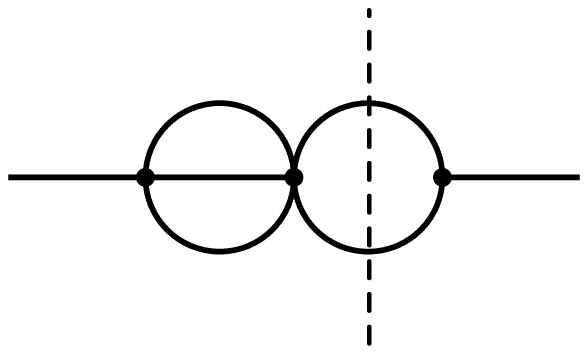}}&
\multicolumn{3}{|c|}{\includegraphics[width=21mm,angle=0]{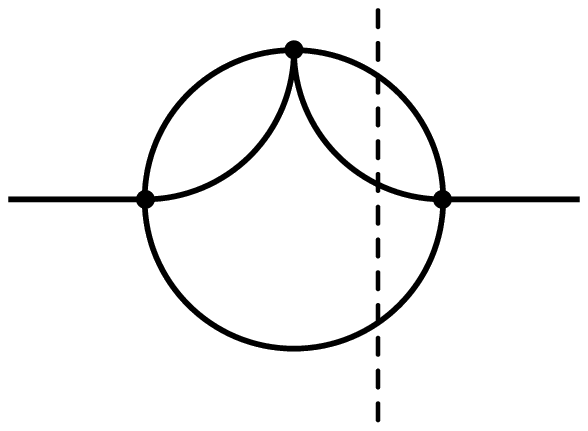}}\\
\multicolumn{2}{|c|}{3L2C1}&\multicolumn{3}{|c|}{3L3C1}\\
\hline
\includegraphics[width=21mm,angle=0]{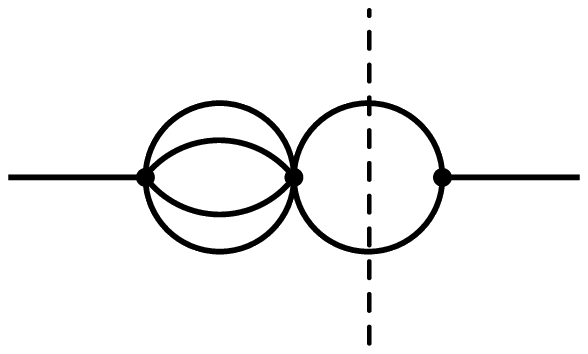} &\includegraphics[width=21mm,angle=0]{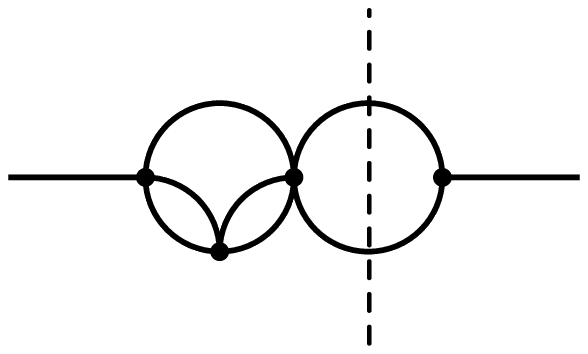} & 
\includegraphics[width=21mm,angle=0]{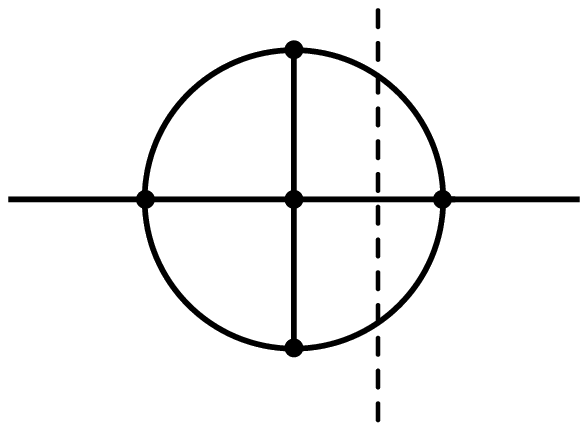} &\includegraphics[width=21mm,angle=0]{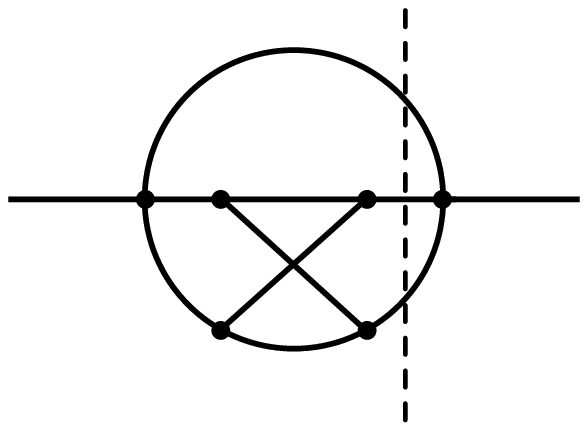}&
\includegraphics[width=21mm,angle=0]{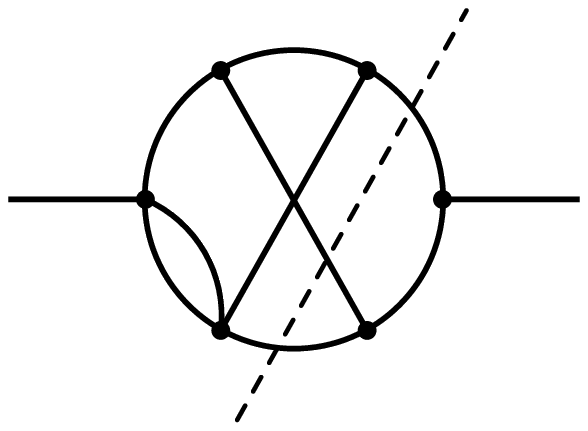}\\
4L2C1&4L2C2&4L3C1&4L3C2&4L3C3\\
\includegraphics[width=21mm,angle=0]{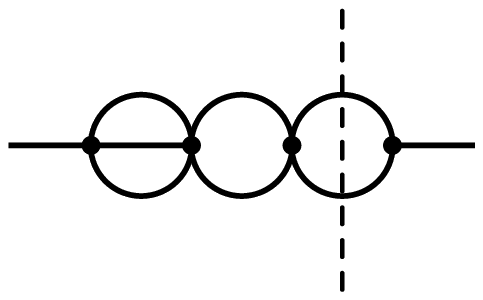} &\includegraphics[width=21mm,angle=0]{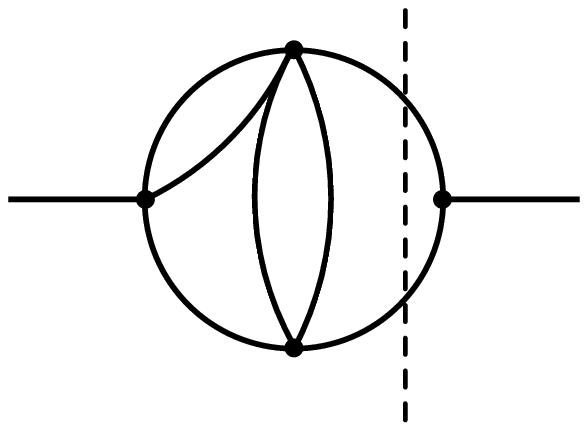} & 
\includegraphics[width=21mm,angle=0]{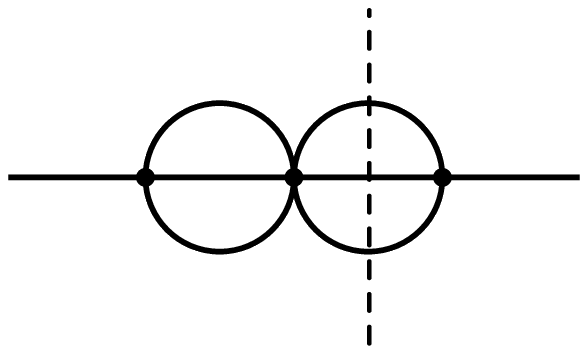} &\includegraphics[width=21mm,angle=0]{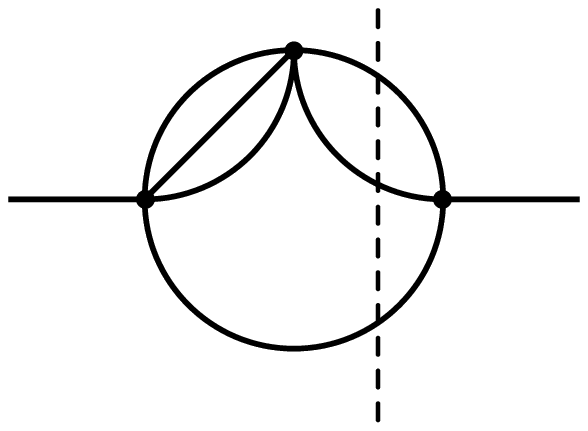}&
\includegraphics[width=21mm,angle=0]{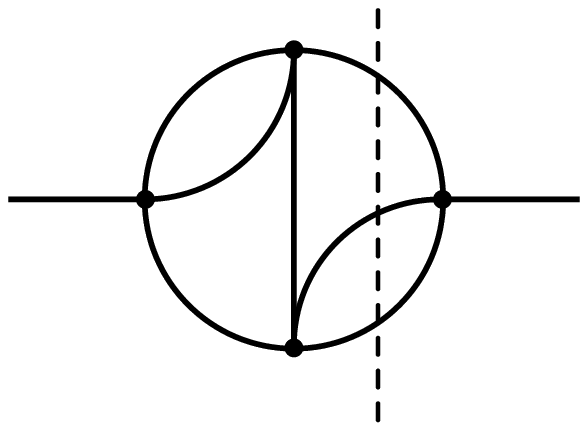}\\
4L2C3&4L2C4&4L3C4&4L3C5&4L3C6\\
\includegraphics[width=21mm,angle=0]{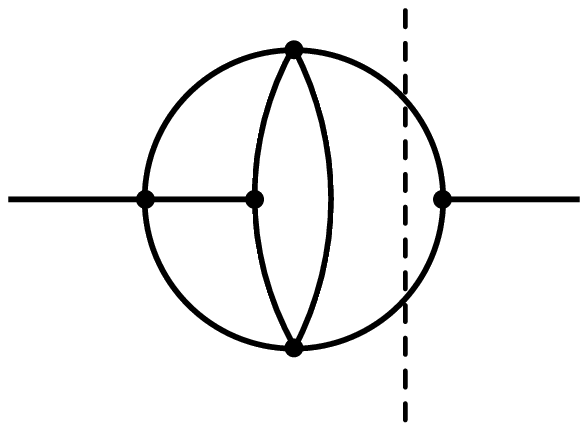} &\includegraphics[width=21mm,angle=0]{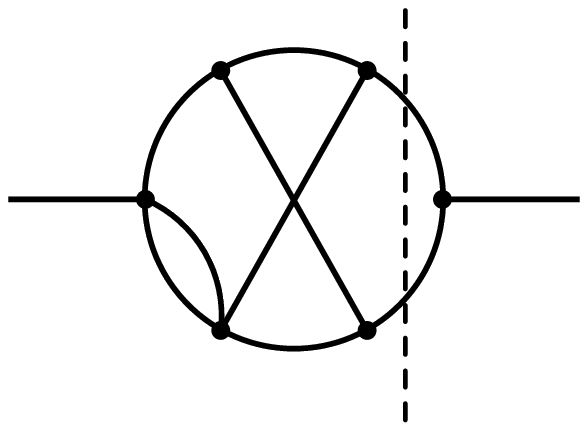} & 
\includegraphics[width=21mm,angle=0]{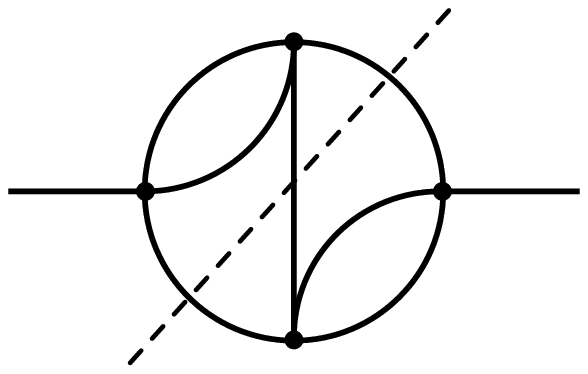} &\includegraphics[width=21mm,angle=0]{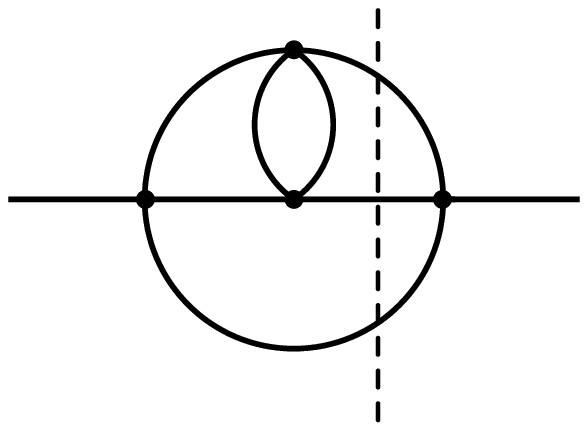}&
\includegraphics[width=21mm,angle=0]{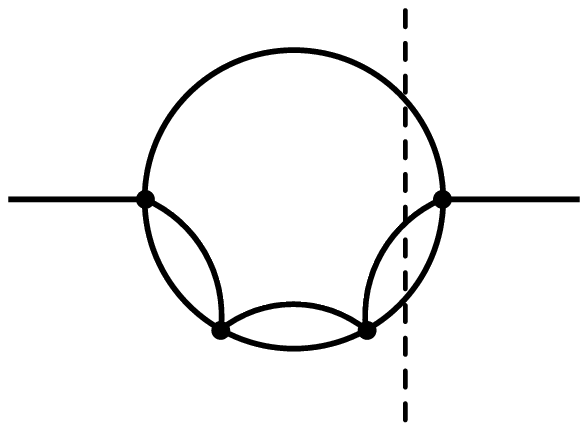}\\
4L2C5&4L2C6&4L3C7&4L3C8&4L3C9\\
\hline
\end{tabular}
\end{center}
\caption{\sf The massless two- and three-particle-cut diagrams used in the course of this work.  \label{tab:masters}}
\end{table}

\label{sec:notation}

In the following, we explain our computation of the
four-particle-cut master integrals in dimensional regularization with
$D=4-2\eps$. The total momentum is $q=p_1+p_2+p_3+p_4$, and we have $p_i^2=0$
for $i=1, \ldots ,4$.  Moreover, all the internal lines are
massless. The momenta are in Minkowski space, and we tacitly
assume that all the propagators below contain an infinitesimal $+i\eta$
with $\eta>0$.  We also define the invariants
\be
s_{ijk\ldots} \equiv (p_i+p_j+p_k+\ldots)^2 \; .
\ee
We therefore have $s_{12}+s_{13}+s_{14}+s_{23}+s_{24}+s_{34}=q^2$ as a
constraint from overall momentum conservation.

Our convention for the loop measure is
\be
\int \! \left[dk\right] \equiv \loopint{k} \; ,
\ee
and we define the prefactor
\be
\ESGamma \equiv \frac{1}{\left(4\pi\right)^{D/2} \Gamma(1-\eps)}\; .
\ee
Note that our definition of $\ESGamma$ is different from the one in Eq.~(4.13)
of Ref.~\cite{GehrmannDeRidder:2003bm}.

As far as integration over the four-particle massless phase space in
$D=4-2\eps$ dimensions is concerned, we closely follow 
Ref.~\cite{GehrmannDeRidder:2003bm}.  The phase space measure reads
\be
dPS_4 = \frac{d^{D-1}p_1}{(2\pi)^{D-1}\, 2E_1} \ldots \frac{d^{D-1}p_4}{(2\pi)^{D-1}\, 2E_4} \, 
        (2\pi)^D \, \delta^{(D)}(q-p_1-p_2-p_3-p_4) \, .
\ee
It can be rewritten in terms of invariants and angular variables according
to
\bea
dPS_4 &=& (2\pi)^{4-3D} \left(q^2\right)^{1-\frac{D}{2}} 2^{1-\frac{D}{2}} \left(-\Delta_4\right)^{\frac{D-5}{2}} 
           \theta(-\Delta_4) \, d\Omega_{D-1}\, d\Omega_{D-2}\, d\Omega_{D-3}\nnb \\
      && \times \delta(q^2-s_{12}-s_{13}-s_{14}-s_{23}-s_{24}-s_{34}) \, 
          ds_{12}\, ds_{13}\, ds_{14}\, ds_{23}\, ds_{24}\, ds_{34} \; , \label{eq:PSinvtrad}
\eea
with the Gram determinant
\be
\Delta_4 = \lambda(s_{12}s_{34},s_{13}s_{24},s_{14}s_{23}) \, , \qquad \lambda(x,y,z) = x^2+y^2+z^2 -2xy-2xz-2yz \; .
\ee

It turns out that integration over angular variables
is trivial in all the cases we encounter here, and we can use
\be
\int \!\! d\Omega_{D} = \frac{2\pi^{D/2}}{\Gamma(D/2)} \; .
\ee
Performing the angular integration, and furthermore applying the steps
explained in Ref.~\cite{GehrmannDeRidder:2003bm} to factorize the phase space
measure, we arrive at
\bea
dPS_4 &=& \frac{2\pi \left(q^2\right)^{2-3\eps}}{(4\pi)^{\frac{3D}{2}} (1-2\eps) \Gamma(1-\eps) \Gamma^2(\frac{1}{2}-\eps)}
          \, dt \, dv \, d\chi \, dz_{1} \, dy_{134} \, dy_{1234} \, \delta(1-y_{1234}) \\
       &&  t^{-\eps} \, (1-t)^{-\eps} \, v^{-\eps} \, (1-v)^{-\eps} \,\chi^{-\frac{1}{2}-\eps} \, 
           (1-\chi)^{-\frac{1}{2}-\eps} \, z_1^{-\eps} \, (1-z_1)^{1-2\eps}\, y_{134}^{1-2\eps} \, (1-y_{134})^{1-2\eps}.\nnb
\eea
All the integration variables $t$,~$v$,~$\chi$,~$z_{1}$,~$y_{134}$, and~$y_{1234}$
run from $0 \ldots 1$ and originate from
\begin{align}
s_{ijk\ldots} &= q^2 \, y_{ijk\ldots} \; , & y_{13} &= (y_{13,\,b}-y_{13,\,a}) \chi + y_{13,\,a} \; , \nnb \\
y_{12} &= \bar y_{134} \, \bar z_1 \, \bar t \; , &  y_{13,\,b/a} &= B \pm \sqrt{B^2-C} \; ,\nnb \\
y_{23} &= \bar y_{134} \, z_1\; , &  B &= y_{134} \left(\bar t \, \bar v + v \, t \, z_1 \right)\; ,\nnb \\
y_{14} &= y_{134} \, \bar z_1 \, v \; , &  C &= y_{134}^2 \left(\bar t \, \bar v - v \, t \, z_1 \right)^2\; ,\nnb \\
y_{24} &= \bar y_{134} \, \bar z_1 \, t \; , & \sqrt{B^2-C} &= 2 \,y_{134} \,\sqrt{t}\,\sqrt{\bar t \,}\,\sqrt{v}\,\sqrt{\bar v}\,\sqrt{z_1} \; , \nnb \\
y_{124} &= \bar z_{1} \left(1- y_{134} \bar v\right) \; , & y_{13,\,b}-y_{13,\,a} &= 2 \sqrt{B^2-C}  \; , \label{eq:substitutions}
\end{align}
where $\bar t = 1-t$, and analogously for all the other variables. The
substitutions~(\ref{eq:substitutions}) should be done in the integrands, too.

\subsection*{A.1~~Results for the four-particle-cut master integrals}\label{sec:results}

We are now in position to present results for the four-particle-cut diagrams
depicted in Fig.~\ref{fig:4PCutMasters}. Normalization factors are
extracted according to
\be \label{normalization4p4}
I_{4L4Ci} = 2\pi \, e^{i\pi\eps} \, \ESGamma^4 \left(q^2\right)^{a_i-4\eps} \, \widetilde I_{4L4Ci} \; ,
\ee
where the $a_i$ follow from dimensional considerations. One finds $a_i =
(2,2,1,-1,0,-1,-1,0)_i$ for $i=1, \ldots ,8$.

\ \\
We start with $I_{4L4C1}$,
\bea
I_{4L4C1} &=& \int \!\! dPS_4  \! \int \! \left[dk\right] \, \frac{1}{k^2 \, (k+p_1+p_2)^2} \nnb \\
    &=& \frac{e^{i\pi\eps} \, \Gamma(\eps) \Gamma^2(1-\eps)}{\left(4\pi\right)^{D/2} \Gamma(2-2\eps)} \, 
        \left(q^2\right)^{-\eps} \! \int \!\! dPS_4 \; y_{12}^{-\eps} \; ,
\eea
which yields
\bea
\widetilde I_{4L4C1} &=& \frac{\Gamma(\eps) \Gamma^9(1-\eps) \Gamma(1-2\eps)\Gamma(2-3\eps)}{
                         \Gamma^2(2-2\eps)\Gamma(3-4\eps)\Gamma(4-5\eps)} \; .
\eea

\ \\
The next integral to consider is $I_{4L4C2}$,
\bea
I_{4L4C2} &=& \int \!\! dPS_4  \! \int \! \left[dk\right] \, \frac{1}{k^2 \, (k+p_1+p_2+p_4)^2} \nnb \\
    &=& \frac{e^{i\pi\eps} \, \Gamma(\eps) \Gamma^2(1-\eps)}{\left(4\pi\right)^{D/2} \Gamma(2-2\eps)} \, \left(q^2\right)^{-\eps} \! \int \!\! dPS_4 \; y_{134}^{-\eps} \; ,
\eea
and we get
\bea
\widetilde I_{4L4C2} &=& \frac{\Gamma(\eps) \Gamma^{10}(1-\eps)\Gamma(2-3\eps)}{\Gamma^2(2-2\eps)\Gamma(3-3\eps)\Gamma(4-5\eps)} \; .
\eea

\ \\
We proceed with $I_{4L4C3}$,
\bea
I_{4L4C3} &=& \int \!\! dPS_4  \! \int \! \left[dk\right] \, \frac{1}{k^2 \, (k+p_1+p_3+p_4)^2 \, (p_1+p_2+p_4)^2} \nnb \\
    &=& \frac{e^{i\pi\eps} \, \Gamma(\eps) \Gamma^2(1-\eps)}{\left(4\pi\right)^{D/2} \Gamma(2-2\eps)} \, \left(q^2\right)^{-1-\eps} \! \int \!\! dPS_4 \; y_{134}^{-\eps} \, y_{124}^{-1} \; ,
\eea
and arrive at
\bea
\widetilde I_{4L4C3} &=& \frac{\Gamma(\eps) \Gamma^{10}(1-\eps)\Gamma(1-2\eps)}{\Gamma^3(2-2\eps)\Gamma(4-5\eps)} \; \pFq{3}{2}{1,1-\eps,2-3\eps}{2-2\eps,4-5\eps}{1}.
\eea
The expansion of $\widetilde I_{4L4C3}$ in $\eps$ is conveniently done with the package {\tt HypExp}~\cite{Huber:2005yg,Huber:2007dx},
\mathindent0cm
\bea
\widetilde I_{4L4C3} &=& \frac{1}{4\eps} + \left(\frac{37}{8}-\frac{\pi^2}{12}\right)+\left(\frac{809}{16}-\frac{35\pi^2}{24}-5\zeta_3\right) \eps
                   +\left(\frac{13677}{32}-\frac{253\pi^2}{16}-\frac{29\pi^4}{144}-71\zeta_3\right) \eps^2 \nnb \\
               && +\left(\frac{198241}{64}-\frac{12995\pi^2}{96}-\frac{3521\pi^4}{1440}-\frac{1287}{2}\zeta_3+\frac{67}{6} \pi^2\zeta_3 -\frac{315}{2}\zeta_5\right) \eps^3
	          +\left(\frac{2597477}{128}\right.\nnb \\
               && \left.-\frac{192175\pi^2}{192}-\frac{17519\pi^4}{960}-\frac{1481\pi^6}{6048}-\frac{19139}{4}\zeta_3+\frac{925}{6} \pi^2\zeta_3+170\zeta_3^2 -2049\zeta_5\right)\eps^4\nnb \\
               && + \, {\mathcal O}(\eps^5) \; .
\eea
\mathindent1cm

\ \\[-2mm]
We now move to $I_{4L4C4}$,
\bea
I_{4L4C4} &=& \int \!\! dPS_4  \! \int \! \left[dk\right] \, \frac{1}{k^2 \, (k+p_1+p_3+p_4)^2 \, (p_1+p_3)^2 \, (p_1+p_2+p_4)^2 \, (p_1+p_2)^2} \nnb \\
    &=& \frac{e^{i\pi\eps} \, \Gamma(\eps) \Gamma^2(1-\eps)}{\left(4\pi\right)^{D/2} \Gamma(2-2\eps)} \, \left(q^2\right)^{-3-\eps} \! \int \!\! dPS_4 \;
    y_{134}^{-\eps} \, y_{124}^{-1} \, y_{13}^{-1} \, y_{12}^{-1} \; ,
\eea
which does not reveal a closed form since we cannot avoid $y_{13}$ in the
integrand. We therefore compute it from the following two-fold Mellin-Barnes
representation~\cite{smirnov,Tausk,Alejandro,smirnovbook,Czakon:2005rk}
\mathindent0cm
\bea
\widetilde I_{4L4C4} &=& \frac{\Gamma(\eps) \Gamma^{6}(1-\eps)\Gamma(-\eps)\Gamma(1-3\eps)}{\Gamma(-2\eps)\Gamma^2(2-2\eps)}
                   \, \MB{c_1}{z_1}\MB{c_2}{z_2} \, \Gamma(z_1+z_2-\eps)\Gamma(-\eps-z_1-z_2)\Gamma(z_1)\nnb\\
               && \times \frac{\Gamma(1-z_1)\Gamma(1-2\eps-z_1)}{\Gamma(2-5\eps-z_1)} \,
	          \frac{\Gamma(-z_2)\Gamma(1+z_2)\Gamma(-1-\eps-z_2)\Gamma(1-\eps+z_2)}{\Gamma(1-3\eps+z_2)\Gamma(-\eps-z_2)} \, .
\eea
\mathindent1cm
The integration contours in the complex plane can be chosen as straight lines
parallel to the imaginary axis. The integral is then 
regulated~\cite{Czakon:2005rk} for $c_1=1/2$,~$c_2=-1/4$, and~$\eps=-7/4$.  We
perform an analytic continuation to $\eps=0$ with the package {\tt
MB.m}~\cite{Czakon:2005rk}, which is also used for numerical cross checks.
The expansion of $\widetilde I_{4L4C4}$ in $\eps$ reads
\bea
\widetilde I_{4L4C4} &=& \frac{1}{4\eps^5}+\frac{1}{\eps^4} + \left(3-\frac{13\pi^2}{24}\right)\frac{1}{\eps^3}+\left(8-\frac{13\pi^2}{6}-\frac{33}{2}\zeta_3\right) \frac{1}{\eps^2}
                   +\left(20-\frac{13\pi^2}{2}-\frac{397\pi^4}{1440}\right. \nnb \\
		   && -66\zeta_3\bigg) \frac{1}{\eps} +\left(48-\frac{52\pi^2}{3}-\frac{397\pi^4}{360}-198\zeta_3+\frac{131}{4} \pi^2\zeta_3 -\frac{687}{2}\zeta_5\right) \nnb \\
               && +\left(112-\frac{130\pi^2}{3}-\frac{397\pi^4}{120}-\frac{24539\pi^6}{60480}-528\zeta_3+131 \pi^2\zeta_3+\frac{897}{2}\zeta_3^2 -1374\zeta_5\right)\eps\nnb \\
               && + \, {\mathcal O}(\eps^2) \; .
\eea

\ \\
The next integral, $I_{4L4C5}$, with
\bea
I_{4L4C5} &=& \int \!\! dPS_4  \! \int \! \left[dk\right] \, \frac{1}{k^2 \, (k+p_4)^2 \, (k+p_1+p_2+p_4)^2 \, (p_2+p_3)^2} \\
    &=& \frac{e^{i\pi\eps} \, \Gamma(1+\eps)\Gamma(-\eps) \Gamma(1-\eps)}{\left(4\pi\right)^{D/2} \Gamma(1-2\eps)} \! \left(q^2\right)^{-2-\eps} \!\! \int \!\! dPS_4
    \! \int\limits_0^1 \! dx \; \frac{1}{\left[y_{12}+x \, y_{14}+x \, y_{24}\right]^{1+\eps}y_{23}} , \nnb
\eea
can again be expressed to all orders in $\eps$. One first integrates over $x$, and finally finds
\bea
\widetilde I_{4L4C5} &=& - \, \frac{\Gamma(\eps) \Gamma^{6}(1-\eps)\Gamma^3(-\eps)}{\Gamma(2-5\eps)\Gamma(2-2\eps)}
                   \left[\frac{\Gamma(1-\eps)}{\Gamma(2-2\eps)} \, \pFq{3}{2}{1,1-\eps,1-2\eps}{1+\eps,2-2\eps}{1}\right. \nnb \\
                && \hspace*{55pt} \left. - \, \frac{\Gamma(1-3\eps)}{(1-3\eps)\Gamma(1-4\eps)} \, \pFq{3}{2}{1,1-\eps,1-3\eps}{1+\eps,2-3\eps}{1}\right] \; .
\eea
The expansion of $\widetilde I_{4L4C5}$ in $\eps$ reads
\bea
\widetilde I_{4L4C5} &=& \frac{2\zeta_3}{\eps^2}+ \left(14\zeta_3+\frac{31\pi^4}{180}\right)\frac{1}{\eps}
                   +\left(78\zeta_3+\frac{217\pi^4}{180}-\frac{20}{3}\pi^2\zeta_3+114\zeta_5\right) \nnb \\
		   && +\left(406\zeta_3+\frac{403\pi^4}{60}-\frac{140}{3}\pi^2\zeta_3+798\zeta_5+\frac{799\pi^6}{7560}-125\zeta_3^2\right) \eps + \, {\mathcal O}(\eps^2) \; .
\eea

\ \\
Also the next integral, $I_{4L4C6}$, with
\bea
I_{4L4C6} &=& \int \!\! dPS_4  \! \int \! \left[dk\right] \, \frac{1}{k^2 \, (k-p_2)^2 \, (k+p_4)^2 \, (k+p_1+p_4)^2 \, (p_1+p_2)^2} \\
    &=& \frac{e^{i\pi\eps} \, \Gamma(2+\eps)\Gamma^2(-\eps)}{\left(4\pi\right)^{D/2} \Gamma(-2\eps)} \! \left(q^2\right)^{-3-\eps} \!\! \int \!\! dPS_4
    \! \int\limits_0^1 \! dx \! \int\limits_0^1 \! dy \; \frac{1}{\left[x \, y_{24}+y \, y_{14}+x y \, y_{12}\right]^{2+\eps}y_{12}} , \nnb
\eea
reveals a closed form which, however, turns out to be more complicated. One first integrates over $x$ and~$y$, and finally finds
\mathindent0cm
{\allowdisplaybreaks
\bea
\widetilde I_{4L4C6} &=& \frac{\Gamma(\eps) \Gamma^{6}(1-\eps)\Gamma^2(-\eps)\Gamma(-1-3\eps)}{\Gamma(1-5\eps)\Gamma(2-2\eps)\Gamma(1-4\eps)}
                   \bigg[ -\frac{3}{2} \, \Gamma(1-2\eps)\Gamma(\eps) -2 \, \Gamma^2(1-2\eps)\Gamma(2\eps)\Gamma(1+\eps)\nnb \\
		   &&  -2\Gamma(1-2\eps)\Gamma(1+\eps) \left(\psi^{(0)}(1-\eps)-\psi^{(0)}(\eps)-\psi^{(0)}(1-4\eps)+2\psi^{(0)}(1-2\eps)+\gamma\right)\nnb \\
		   && -4\Gamma(-\eps) \, \pFq{3}{2}{1,-\eps,-\eps}{1+\eps,1-\eps}{1} 
		      - \frac{4\Gamma^2(-2\eps)}{\Gamma(-3\eps)} \, \pFq{3}{2}{-\eps,-\eps,-\eps}{-3\eps,1-\eps}{1} \nnb \\
		   &&  + \frac{\Gamma^2(1-\eps)\Gamma(1-4\eps)}{(1+\eps)^2\Gamma(1-3\eps)\Gamma(-2\eps)}\, \pFq{4}{3}{1,1-\eps,1-\eps,1+\eps}{2+\eps,2+\eps,1-3\eps}{1} \nnb \\
		   &&  - \, \frac{\Gamma^2(1-2\eps)\Gamma(1+\eps)}{\Gamma(-2\eps)}\, \pFq{4}{3}{1,1,1-2\eps,1-2\eps}{2,2,1-4\eps}{1}\bigg] \; ,
\eea
where $\psi^{(0)}(z) = \f{d}{dz}\ln\Gamma(z)$. 
The expansion of $\widetilde I_{4L4C6}$ in $\eps$ reads
\bea
\widetilde I_{4L4C6} &=& \frac{5}{6\eps^5}-\frac{5}{6\eps^4} + \left(\frac{35}{6}-\frac{79\pi^2}{36}\right)\frac{1}{\eps^3}+\left(-\frac{65}{6}+\frac{79\pi^2}{36}-58\zeta_3\right) \frac{1}{\eps^2}
                   +\left(\frac{275}{6}-\frac{553\pi^2}{36}\right. \nnb \\
		   && \left.+\frac{643\pi^4}{2160}+58\zeta_3\right) \frac{1}{\eps}
		     +\left(-\frac{665}{6}+\frac{1027\pi^2}{36}-\frac{643\pi^4}{2160}-406\zeta_3+\frac{1301}{9} \pi^2\zeta_3 -\frac{2590}{3}\zeta_5\right) \nnb \\
               && +\left(\frac{2315}{6}-\frac{4345\pi^2}{36}+\frac{4501\pi^4}{2160}+\frac{63229\pi^6}{272160}+754\zeta_3-\frac{1301}{9} \pi^2\zeta_3+1884\zeta_3^2 +\frac{2590}{3}\zeta_5\right)\eps\nnb \\
               && + \, {\mathcal O}(\eps^2) \; .
\eea}

\ \\
The next integral, $I_{4L4C7}$, has not been necessary for the actual
calculation of $\widetilde{G}^{(2)}_{17}$ and $\widetilde{G}^{(2)}_{27}$ because
it stems from diagrams where the charm quark loop is cut.  However, we still
give the result, as it is the most complicated integral, and might 
be useful for future computations of other interferences. The difficulty
is due to the fact that one cannot avoid $y_{13}$ in the integrand, and the
resulting Mellin-Barnes representation is four-dimensional. Starting from
\bea
I_{4L4C7} &=& \int \!\! dPS_4  \! \int \! \left[dk\right] \, \frac{1}{k^2 \, (k-p_1)^2 \, (k+p_2+p_3+p_4)^2 \, (k+p_3+p_4)^2 \, (p_1+p_2+p_3)^2} \nnb \\
    &=& \frac{e^{i\pi\eps} \, \Gamma(2+\eps)\Gamma^2(-\eps)}{\left(4\pi\right)^{D/2} \Gamma(-2\eps)} \! \left(q^2\right)^{-3-\eps} \nnb \\
    && \times \int \!\! dPS_4 \! \int\limits_0^1 \! dx \! \int\limits_0^1 \! dy \; \frac{1}{\left[y_{34}+x \, (y_{13}+y_{14})+y \, (y_{23}+y_{24}) + x y \, y_{12}\right]^{2+\eps}y_{123}} \; ,
\eea
we first integrate over $x$ and~$y$, and find the following Mellin-Barnes representation.
{\allowdisplaybreaks
\bea
\widetilde I_{4L4C7} &=& \frac{\Gamma(\eps) \Gamma^{5}(1-\eps)\Gamma(-\eps)\Gamma(1-3\eps)}{\Gamma(-2\eps)\Gamma(2-2\eps)} \, \MB{c_1}{z_1}\MB{c_2}{z_2}\nnb \\
                && \times  \frac{\Gamma(-\eps-z_1)\Gamma(1-\eps+z_1)\Gamma(1-3\eps+z_1-z_2)\Gamma(1-2\eps-z_2)\Gamma(-\eps-z_1+z_2)}{\Gamma(1-z_2-3\eps)\Gamma(1-z_2-4\eps)\Gamma(1+z_2-\eps)} \nnb\\
               && \times \frac{\Gamma(-z_1)\Gamma(1+z_1)\Gamma(-z_2)\Gamma(1+z_2)\Gamma(-z_2-\eps)\Gamma(z_2-\eps)}{\Gamma(1-z_1-3\eps)\Gamma(2+z_1-3\eps)} \nnb \\
                &&- \, \frac{2\Gamma(\eps) \Gamma^{5}(1-\eps)\Gamma(-\eps)\Gamma^2(1-3\eps)}{\Gamma(1-5\eps)\Gamma(1-2\eps)\Gamma(-2\eps)\Gamma(2-2\eps)} \, \MB{c_1}{z_1}\MB{c_2}{z_2}\MB{c_3}{z_3}\nnb \\
                && \times \frac{\Gamma(-z_1)\Gamma(1+z_1-z_3)\Gamma(-z_2)\Gamma(1+z_2)\Gamma(-z_1+z_3-\eps)\Gamma(z_2-\eps)\Gamma(-z_2-z_3-\eps)}{\Gamma(1-z_1+z_2+z_3-4\eps)\Gamma(2+z_1-z_3-3\eps)}\nnb\\
               && \times \frac{\Gamma(z_3)\Gamma(1-4\eps+z_2+z_3)\Gamma(1-2\eps+z_1)\Gamma(-z_1+z_2+z_3-\eps)\Gamma(1-\eps+z_1-z_3)}{\Gamma(1+z_3-3\eps)\Gamma(1+z_2-\eps)} \nnb \\
                &&+ \frac{\Gamma(\eps) \Gamma^{5}(1-\eps)\Gamma(-\eps)\Gamma(1-3\eps)}{\Gamma(1-5\eps)\Gamma(1-2\eps)\Gamma(-2\eps)\Gamma(2-2\eps)}
		    \, \MB{c_1}{z_1}\MB{c_2}{z_2}\MB{c_3}{z_3}\MB{c_4}{z_4} \nnb \\
                && \times \frac{\Gamma(-z_3)\Gamma(z_3-z_1)\Gamma(-z_2)\Gamma(1+z_2)\Gamma(-z_4)\Gamma(1+z_1+z_4)\Gamma(z_2-\eps)}{\Gamma(1-z_1+z_2+z_3-z_4-4\eps)\Gamma(1+z_1-z_3-\eps)}\nnb\\
               && \times \frac{\Gamma(1-\eps+z_1)\Gamma(z_1-z_2-z_3-\eps)\Gamma(-z_1+z_3-z_4-\eps)\Gamma(-z_1+z_2+z_3-z_4-\eps)}{\Gamma(1+z_3-\eps)\Gamma(1+z_2-\eps)} \nnb \\
               && \times \Gamma(-z_1-2\eps)\Gamma(1+z_2+z_3-2\eps)\Gamma(1+z_1-z_3+z_4-\eps) \; .
\eea}
\mathindent1cm
The expansion of $\widetilde I_{4L4C7}$ in $\eps$ reads
\bea
\widetilde I_{4L4C7} &=& -\frac{2\pi^4}{45\eps} + \left(-\frac{16\pi^4}{45}+2\pi^2\zeta_3-58 \zeta_5\right) \nnb \\
		   && +\left(-\frac{104\pi^4}{45}+16\pi^2\zeta_3-464 \zeta_5+84\zeta_3^2-\frac{1289\pi^6}{5670}\right) \eps + \, {\mathcal O}(\eps^2) \; . \label{eq:N7texp}
\eea
We have also derived an alternative, seven-fold, Mellin-Barnes representation for
$\widetilde I_{4L4C7}$ and used it to confirm~(\ref{eq:N7texp}) numerically with the
help of the code {\tt MB.m}~\cite{Czakon:2005rk}.

\ \\
The last integral, $I_{4L4C8}$, reads
\bea
I_{4L4C8} &=& \int \!\! dPS_4  \! \int \! \left[dk\right] \, \frac{1}{k^2 \, (k+p_1+p_2+p_4)^2 \, (k+p_1+p_2)^2 \, (p_1+p_3+p_4)^2} \\
    &=& \frac{e^{i\pi\eps} \, \Gamma(1+\eps)\Gamma(-\eps) \Gamma(1-\eps)}{\left(4\pi\right)^{D/2} \Gamma(1-2\eps)} \! \left(q^2\right)^{-2-\eps} \!\! \int \!\! dPS_4
    \!\! \int\limits_0^1 \! dx \; \frac{1}{\left[y_{12}+x \, y_{14}+x \, y_{24}\right]^{1+\eps}y_{134}} .\nnb
\eea
Again, one first integrates over $x$, and finally finds an expression involving
a one-dimensional Feynman parameter integral
{\allowdisplaybreaks
\bea
\widetilde I_{4L4C8} &=&\frac{\Gamma(1-3 \eps) \Gamma(1-2
   \eps) \Gamma^4 (1-\eps) \Gamma^4 (-\eps) \Gamma(2 \eps) \Gamma^3 (1+\eps)}{\Gamma(2-5 \eps)
   \Gamma(2-4 \eps) \Gamma(2-2 \eps) \Gamma(3 \eps)}\nnb \\
&&
+\frac{\Gamma^2 (1-3 \eps) \Gamma(1-2 \eps) \Gamma^4
   (1-\eps) \Gamma^3 (-\eps) \Gamma^2 (1+\eps) \Gamma(2 \eps)}{\Gamma(2-5 \eps) \Gamma(2-4
   \eps) \Gamma(2-2 \eps)} \nnb \\
&&
-\frac{\Gamma(1-3 \eps) \Gamma^5 (1-\eps) \Gamma^4 (-\eps) \Gamma(1+\eps) }{2 \, \Gamma(2-5 \eps) \Gamma(2-4 \eps) \Gamma(2-2
   \eps)} \, \pFq{3}{2}{1,1-\eps,2 \eps}{1+\eps,1+2 \eps}{1}\nnb \\
&&
-\frac{ \Gamma(1-3 \eps) \Gamma^7(1-\eps) \Gamma^2 (-\eps) \Gamma(\eps) }{2 \, \Gamma(3-5 \eps) \Gamma^2 (2-2 \eps) \Gamma(-2 \eps)}\nnb \\
&& \times \int\limits_{0}^{1} \!\! dt \, \, t^{1-2 \eps} \, \left(1-t\right)^{-\eps}
\, \pFq{2}{1}{1,2-4 \eps}{3-5 \eps}{t} \; \pFq{2}{1}{1,1-\eps}{2-2\eps}{t} \; .
\eea}
The expansion of $\widetilde I_{4L4C8}$ in $\eps$ reads
\bea
\widetilde I_{4L4C8} &=& -\frac{\zeta_3}{\eps}+\left(-11 \zeta_3-\frac{19 \pi ^4}{360}\right)
                   + \left(-83 \zeta_3+\frac{23 \pi ^2 \zeta_3}{6}-36 \zeta_5-\frac{209 \pi ^4}{360}\right)\eps \nnb \\
		   && + \left(-535 \zeta_3+\frac{253 \pi ^2 \zeta_3}{6}+70 \zeta_3^2-396 \zeta_5-\frac{1577 \pi ^4}{360}+\frac{13 \pi ^6}{378}\right) \eps^2+ \, {\mathcal O}(\eps^3) \; .
\eea

\subsection*{A.2~~Results for the three-particle-cut master integrals}\label{sec:results3pcut}

In this section, we describe our computation of the three-particle-cut diagrams
4L3C1, 4L3C2 and 4L3C3.  Similarly to Eq.~(\ref{normalization4p4}), we extract the normalization
factors according to
\be
I_{4L3Ci} = 2\pi \, e^{2\pi i\eps} \, \ESGamma^4 \left(q^2\right)^{b_i-4\eps} \, \widetilde I_{4L3Ci} \; ,
\ee
where the $b_i$ again follow from dimensional considerations. One finds $b_1 = 0$ and $b_2 = -1$.
For 4L3C3, we have used a different method, as explained below.

The kinematics and the phase space measure are much simpler in the
three-particle case, compared to the four-particle one. The total momentum is
$q=p_1+p_2+p_3$, and we have $p_i^2=0$ for $i=1, \ldots ,3$.  We define the
invariants
\be
s_{ijk\ldots} \equiv (p_i+p_j+p_k+\ldots)^2 
\ee
as before, and have $s_{12}+s_{13}+s_{23}=q^2$ as a constraint from overall
momentum conservation. The phase space measure
\be
dPS_3 = \frac{d^{D-1}p_1}{(2\pi)^{D-1}\, 2E_1} \ldots \frac{d^{D-1}p_3}{(2\pi)^{D-1}\, 2E_3} \, (2\pi)^D \, \delta^{(D)}(q-p_1-p_2-p_3)
\ee
is again taken over from Ref.~\cite{GehrmannDeRidder:2003bm}. After integration over
angular variables one finds
\bea
dPS_3 &=& \frac{2\pi \, \ESGamma^2 \, \Gamma^2(1-\eps) \, \left(q^2\right)^{1-2\eps}}{\Gamma(2-2\eps)}
          \, dy_{12} \, dy_{13} \, dy_{23} \; y_{12}^{-\eps} \, y_{13}^{-\eps} \, y_{23}^{-\eps} \,\delta(1-y_{12}-y_{13}-y_{23}) . \nnb
\eea
The integration variables $y_{12}$,~$y_{13}$, and~$y_{23}$ run from $0 \ldots
1$, and originate from $s_{ij} = q^2 \, y_{ij}$. The latter substitutions have to
be made in the integrands, as well.

\ \\
Our first three-particle-cut integral $I_{4L3C1}$ reads
\mathindent0cm
\bea
I_{4L3C1} &=& \int \!\! dPS_3  \! \int \! \left[dk_1\right] \! \int \! \left[dk_2\right] \, \frac{1}{k_1^2 \, (k_1+p_1)^2 \, k_2^2 \,(k_2+p_3)^2 \, (k_1+k_2-p_2)^2} \\
    &=& \frac{e^{2\pi i\eps} \, \ESGamma^2 \, \Gamma^2(-\eps) \Gamma^3(1-\eps)\Gamma(1+2\eps)}{\Gamma(1-3\eps)} \! \left(q^2\right)^{-1-2\eps} \!\! \int \!\! dPS_3
    \! \int\limits_0^1 \! dx \! \int\limits_0^1 \! dy \; \frac{1}{\left[x \, y_{12}+x \, y \,y_{13}+y \, y_{23}\right]^{1+2\eps}} \, . \nnb
\eea
It can be expressed in a closed form valid to all orders in $\eps$. One first integrates over $x$, and finally finds
{\allowdisplaybreaks
\bea
\widetilde I_{4L3C1}&=&-\frac{3 \, \Gamma(1-2 \eps) \Gamma(-3
   \eps) \Gamma^2(-\eps) \Gamma(\eps) \Gamma(2 \eps) \Gamma(2 \eps+1) \Gamma^5(1-\eps)}{2 \, \Gamma(2-5
   \eps) \Gamma(2-2 \eps)} \\
&+&\frac{\Gamma^4(-\eps) \Gamma(2 \eps) \Gamma^5(1-\eps) }{(2 \eps-1)^2 \, \Gamma(2-5 \eps) \Gamma(-2 \eps)} \; \pFq{3}{2}{1,1-\eps,1-2 \eps}{2-2
   \eps,1+\eps}{1}\nnb \\
&+&\frac{\Gamma^2(1-2 \eps) \Gamma^4
   (-\eps) \Gamma(1+\eps) \Gamma(2 \eps) \Gamma^4(1-\eps) }{\Gamma(2-4 \eps) \Gamma(1-3 \eps) \Gamma(2-2 \eps)}\; \pFq{3}{2}{\eps,1-2 \eps,1-2 \eps}{2-4
   \eps,1+\eps}{1}\nnb \\
&-&\frac{\Gamma(1-2 \eps) \Gamma^5(-\eps) \Gamma(2 \eps) \Gamma^5(1-\eps) }{4 \, \Gamma(1-3 \eps) \Gamma(2-3 \eps) \Gamma(2-2 \eps) \Gamma(-2
   \eps)}\; \pFq{4}{3}{1,2\eps,1-\eps,1-\eps}{2-3 \eps,1+\eps,1+2 \eps}{1} \; . \nnb
\eea}
\mathindent1cm
The expansion of $\widetilde I_{4L3C1}$ in $\eps$ reads
\bea
\widetilde I_{4L3C1} &=& \frac{2\zeta_3}{\eps^2}+ \left(14\zeta_3+\frac{\pi^4}{9}\right)\frac{1}{\eps}
                   +\left(78\zeta_3+\frac{7\pi^4}{9}-6\pi^2\zeta_3+78\zeta_5\right) \nnb \\
		   && +\left(406\zeta_3+\frac{13\pi^4}{3}-42\pi^2\zeta_3+546\zeta_5+\frac{5\pi^6}{63}-140\zeta_3^2\right) \eps + \, {\mathcal O}(\eps^2) \; .
\eea

\ \\
The next three-particle-cut integral is $I_{4L3C2}$,
\bea
I_{4L3C2} &=& \int \!\! dPS_3  \! \int \! \left[dk_1\right] \! \int \! \left[dk_2\right] \, \frac{1}{(k_1+p_1+p_2)^2 \, k_1^2 \, (k_1-k_2+p_1)^2 \,(k_1-k_2)^2 \, (k_2+p_2)^2 \, k_2^2} \, .\nnb \\
\eea
Despite the fact that $p_3$ does not appear in the integrand, the result of
the integral is quite lengthy. In the end, we find the following expression 
that involves a one-dimensional Feynman parameter integral:
\mathindent0cm
{\allowdisplaybreaks
\bea
\widetilde I_{4L3C2}&=&\frac{\Gamma(-3\eps-1)\Gamma(-\eps) \Gamma(\eps) \Gamma^6(1-\eps)\Gamma^3(-2\eps)\Gamma^2(1+2\eps)}
{\Gamma(1-5 \eps)\Gamma(2-2 \eps)\Gamma(-4\eps)}\nnb \\
&&
+\frac{\Gamma(-3\eps-1)\Gamma^2(-\eps) \Gamma(2\eps) \Gamma^7(1-\eps)\Gamma(-2\eps)}
{\Gamma(1-5 \eps)\Gamma(2-2 \eps)\Gamma(1-4\eps)\Gamma(-3\eps)\Gamma(2+2 \eps)}
\, \pFq{3}{2}{1,1,1-\eps}{1-4\eps,2+2\eps}{1}\nnb \\
&&
-\frac{\Gamma(-3\eps-1)\Gamma^3(-\eps) \Gamma^2(1+2\eps) \Gamma^7(1-\eps)\Gamma^2(-2\eps)}
{\Gamma(1-5 \eps)\Gamma(2-2 \eps)\Gamma(1-2\eps)\Gamma^2(-3\eps)\Gamma(2+2 \eps)}
\, \pFq{3}{2}{1,1,1-\eps}{1-2\eps,2+2\eps}{1}\nnb \\
&&
+ \frac{\Gamma(-3\eps-1)\Gamma^2(-\eps) \Gamma(2\eps) \Gamma^6(1-\eps)\Gamma(-2\eps)}
{\Gamma(1-5 \eps)\Gamma(2-2 \eps)\Gamma^2(-3\eps)\Gamma(2+2 \eps)}
\int\limits_{0}^{1} \!\! dt \, \, t^{-\eps} \, \left(1-t\right)^{-3\eps-1} \nnb \\
&& \times \, \left[\pFq{2}{1}{-2 \eps,-2 \eps}{1-2 \eps}{1-t}-1\right] \; \pFq{2}{1}{1,1}{2+2\eps}{t}\nnb\\
&&
- \frac{2\, \Gamma(-3\eps-1)\Gamma^2(-\eps) \Gamma^2(1+2\eps) \Gamma^6(1-\eps)\Gamma^2(-2\eps)}
{\Gamma(1-5 \eps)\Gamma(2-2 \eps)\Gamma^2(-3\eps)\Gamma(2+2 \eps)}
\int\limits_{0}^{1} \!\! dt \, \, t^{-\eps} \, \left(1-t\right)^{-\eps-1} \nnb \\
&& \times \, \left[\pFq{2}{1}{-2 \eps,-2 \eps}{-3 \eps}{1-t}-1\right] \; \pFq{2}{1}{1,1}{2+2\eps}{t}\nnb\\
&&
+ \frac{\Gamma(-3\eps-1)\Gamma^2(-\eps) \Gamma^2(1+2\eps) \Gamma^6(1-\eps)\Gamma(-2\eps)}
{\Gamma(1-5 \eps)\Gamma(2-2 \eps)\Gamma^2(-3\eps)\Gamma^2(2+2 \eps)}
\int\limits_{0}^{1} \!\! dt \, \, t^{1+\eps} \, \left(1-t\right)^{-3\eps-1} \nnb \\
&& \times \, \left[\pFq{2}{1}{-2 \eps,-2 \eps}{-3 \eps}{1-t}-1\right] \;
\left[\pFq{2}{1}{1,1}{2+2\eps}{t}\right]^2 \; .
\eea}
The expansion of $\widetilde I_{4L3C2}$ in $\eps$ reads
\bea
\widetilde I_{4L3C2} &=& \frac{1}{3\eps^5}-\frac{1}{3\eps^4} + \left(\frac{7}{3}-\frac{13\pi^2}{18}\right)\frac{1}{\eps^3}+\left(\frac{13\pi^2}{18}-\frac{13}{3}-\frac{61}{3}\zeta_3\right) \frac{1}{\eps^2}
                   +\left(\frac{55}{3}-\frac{91\pi^2}{18}-\frac{11\pi^4}{180}\right. \nnb \\
		   && +\frac{61}{3}\zeta_3\bigg) \frac{1}{\eps} +\left(\frac{169\pi^2}{18}-\frac{133}{3}+\frac{11\pi^4}{180}-\frac{427}{3}\zeta_3+\frac{353}{9} \pi^2\zeta_3 -233\zeta_5\right) \nnb \\
               && +\left(\frac{463}{3}-\frac{715\pi^2}{18}-\frac{77\pi^4}{180}+\frac{17\pi^6}{140}+\frac{793}{3}\zeta_3-\frac{353}{9} \pi^2\zeta_3+\frac{1763}{3}\zeta_3^2 +233\zeta_5\right)\eps\nnb \\
               && + \, {\mathcal O}(\eps^2) \; .
\eea
\mathindent1cm

\ \\
For the last integral $I_{4L3C3}$, we employ a different approach. Due to the
structure of the integrand, it is not possible to find a regulated
Mellin-Barnes representation. Therefore, we begin with evaluating an 
integral $I_{4L3C3'}$ defined as
\bea
I_{4L3C3'} &=& \int \!\! dPS_3  \! \int \! \left[dk_1\right] \! \int \! \left[dk_2\right] \, \frac{1}{\left[(k_1+k_2)^2\right]^2 \, (k_2+p_2)^2 \, k_1^2 \,(k_1+p_3)^2 \, (k_1+p_1+p_3)^2 \, s_{12}} \nnb \\
    &=& \frac{e^{2\pi i\eps} \, \ESGamma^2 \, \Gamma^2(-\eps) \Gamma^3(1-\eps)\Gamma(2+2\eps)\Gamma(-2\eps)}{\Gamma(1-2\eps)\Gamma(-3\eps)} \! \left(q^2\right)^{-3-2\eps} \nnb \\
    && \times \! \int \!\! dPS_3 \! \int\limits_0^1 \! dx \! \int\limits_0^1 \! dy \; \frac{y^{\eps}}{\left[x \, y \, y_{12}+x \,y_{13}+y \, y_{23}\right]^{2+2\eps} \, y_{12}} \, .
\eea
Again, we extract the normalization factor according to 
\be
I_{4L3C3'} = 2\pi \, e^{2\pi i\eps} \, \ESGamma^4 \left(q^2\right)^{-2-4\eps} \, \widetilde I_{4L3C3'} \; ,
\ee
The above quantity can be expressed in terms of a one-dimensional Feynman parameter integral as follows:
{\allowdisplaybreaks
\bea
\widetilde I_{4L3C3'} &=&\frac{3\, \Gamma^4(-\eps) \Gamma(2\eps) \Gamma^6 (1-\eps)}{4\, \Gamma^2(1-3 \eps)\Gamma(2-2 \eps)}
-\frac{5\,\Gamma^2 (1-2 \eps) \Gamma^5(1-\eps) \Gamma^3 (-\eps) \Gamma^2 (2\eps) \Gamma(1+ \eps)}{2\, \Gamma(1-5 \eps) \Gamma(2-2 \eps)} \nnb \\
&&
+\frac{5 \, \Gamma^4 (1-\eps) \Gamma^5 (-\eps) \Gamma(1+2\eps) }{2 \, \Gamma(1-5 \eps) \Gamma(2-2\eps)} \, \pFq{3}{2}{1,-\eps,-\eps}{1-\eps,1+\eps}{1}\nnb \\
&&
+\frac{ 3 \, \Gamma^6(1-\eps) \Gamma^3 (-\eps) \Gamma(2\eps) }{2 \, \Gamma(1-3 \eps) \Gamma(1-2 \eps) \Gamma(2-2 \eps)} \int\limits_{0}^{1} \!\! dt \, \, t^{-2 \eps} \, \left(1-t\right)^{-2\eps-1} \nnb \\
&& \times \, \left[\pFq{2}{1}{1,-5 \eps}{1-2 \eps}{1-t}-1\right] \; \pFq{2}{1}{-\eps,-2\eps}{1-2\eps}{t} \; .
\eea}
The expansion of $\widetilde I_{4L3C3'}$ in $\eps$ reads
\bea
\widetilde I_{4L3C3'} &=& \frac{1}{(1-2\eps)} \left[-\frac{3}{2\eps^5}+\frac{37\pi^2}{12\eps^3}+\frac{100\zeta_3}{\eps^2}+\frac{149\pi^4}{80\eps}+1727\zeta_5-\frac{505}{3}\pi^2\zeta_3\right. \nnb \\
&&\hspace*{52pt} \left.+\left(\frac{186493\pi^6}{90720}-2680\zeta_3^2\right) \, \eps + {\mathcal O}(\eps^2)\right] \; .
\eea
The original integral $I_{4L3C3}$ can then be obtained by relating it to
$I_{4L3C3'}$ with the help of integration-by-parts identities.

\newappendix{Appendix~B:~~ Relation to Ref.~\cite{Buras:2002tp}}
\def\theequation{B.\arabic{equation}}

Several decimal numbers in subsection~\ref{subsec:bare} can be related to the
quantities encountered in Ref.~\cite{Buras:2002tp} as follows.  In the finite
part of $\hat{G}_{47}^{(1)\rm bare}$ in Eq.~(\ref{lower.bare}), we have
\bea
43.76456245573869 &=& Y_1 ~\equiv~
\f{19039}{486} + \f{11}{27} \pi^2 - \f{\pi}{9\sqrt{3}} - \f{16}{27} X_b + \f16 {\rm Re}[a(1)-2b(1)],\nnb\\[2mm]
0.04680853247986 &=& Y_2 ~\equiv~ 2 {\rm Re}\,b(1) - \f{4}{243},
\eea
where 
\bea
X_b &=& -\f98 - \f{\pi^2}{5} - \f23\zeta_3 + \f{1}{10}\,\psi^{(1)}\!\left(\f16\right),\nnb\\[1mm]
{\rm Re}\,a(1) &=& \f{16}{3} +\f{164}{405}\pi^2 -\f{16}{9}\zeta_3 - \f{300\pi + 64\pi^3}{135\sqrt{3}}
+ \f{32\pi\sqrt{3}-72}{405}\,\psi^{(1)}\!\left(\f16\right),\nnb\\[1mm]
{\rm Re}\,b(1) &=& \f{320}{81} + \f{632}{1215} \pi^2 
- \f{4 \pi}{3 \sqrt{3}}- \f{8}{45}\,\psi^{(1)}\!\left(\f16\right),
\eea
and
\be
\psi^{(1)}(z) = \f{d^2}{dz^2} \ln \Gamma(z).
\ee
The above exact expressions for $X_b$ and ${\rm Re}\,a(1)$ are new. They come
from the three-fold Feynman parameter integrals in Eqs.~(3.2) and (3.3) of
Ref.~\cite{Buras:2002tp}.

In the $\f{1}{\eps}$-part of $\widetilde{G}_{27}^{(2)\rm bare}$ in Eq.~(\ref{bare.nnlo}), we have
\bea
-67.66077706444119 &=& -\f23 Y_1 
-\f{103762}{2187} + \f{44}{27} \pi^2 - \f{160}{27} \zeta_3,\nnb\\[1mm]
5.17409838118169 &=& -\f23 Y_2 + \f{11384}{2187}.
\eea
Finally, in the coefficients multiplying $\ln(\mu/m_b)$ in Eq.~(\ref{renormalized}), we have
\bea
1.0460332197 &=& -\f43 Y_1 -\f{37708}{729} + \f{304}{27} \pi^2,\nnb\\[1mm]
9.6604967166 &=& -\f43 Y_2 + \f{7088}{729}.
\eea

\newappendix{Appendix~C:~~ NLO results of relevance for Section~\ref{sec:mcdep}}
\def\theequation{C.\arabic{equation}}

The NLO quantities $K_{ij}^{(1)}$ that occur in Eq.~(\ref{k27nnlo}) are given by
\bea 
K_{27}^{(1)} &=&  -6 K_{17}^{(1)} ~=~  
{\rm Re}\,r_2^{(1)} - \f{208}{81} L_b + 2 \phi_{27}^{(1)}(\delta),\nnb\\
K_{47}^{(1)} &=&  {\rm Re}\,r_4^{(1)} + \f{76}{243} L_b + 2 \phi_{47}^{(1)}(\delta),\nnb\\
K_{77}^{(1)} &=& -\f{182}{9} + \f{8}{9}\pi^2 - \f{32}{3} L_b + 4\,\phi_{77}^{(1)}(\delta),\nnb\\
K_{78}^{(1)} &=&  \f{44}{9} - \f{8}{27}\pi^2 +\f{16}{9} L_b  
                  + 2\,\phi_{78}^{(1)}(\delta),
\eea
where $r_2^{(1)}$ and $r_4^{(1)}$ can be found in Eq.~(3.1) of Ref.~\cite{Buras:2002tp}.
The function $\phi^{(1)}_{27}$ has been already given in Eq.~(\ref{phi27}) here.
The remaining ones read
\bea
\phi^{(1)}_{77} &=& -\f{2}{3} \ln^2 \delta -\f{7}{3} \ln \delta - \f{31}{9} + \f{10}{3} \delta 
+ \f{1}{3} \delta^2 -\f{2}{9} \delta^3 + \f{1}{3} \delta ( \delta - 4 ) \ln \delta,\nnb\\[1mm]
\phi^{(1)}_{78} &=& \f{8}{9} \left[ {\rm Li}_2(1-\delta) - \f{1}{6}\pi^2 - \delta \ln \delta + \f{9}{4} \delta 
                    - \f{1}{4} \delta^2 + \f{1}{12} \delta^3 \right],\nnb\\[1mm]
\phi_{47}^{(1)}(\delta) &=& \phi_{47}^{(1)A}(\delta) + \phi_{47}^{(1)B}(\delta),
\eea
where\footnote{
$\;$Eq.~(3.12) of Ref.~\cite{Misiak:2006ab} gives $\phi_{47}^{(1)A}$ only, and
contains a misprint in the coefficient at~ $\lim_{m_c \to m_b}$.}
\bea
\phi_{47}^{(1)A}(\delta) &=& \f{1}{54} \pi \left( 3 \sqrt{3} - \pi \right) 
  + \f{1}{81}\delta^3 - \f{25}{108} \delta^2 + \f{5}{54} \delta
  + \f{2}{9} \left( \delta^2 + 2\delta + 3 \right) \arctan^2\sqrt{\f{1-\delta}{3+\delta}}\nnb\\[1mm]
&-& \f{1}{3} \left( \delta^2 + 4\delta + 3 \right) \sqrt{\f{1-\delta}{3+\delta}}\, 
    \arctan\sqrt{\f{1-\delta}{3+\delta}},\nnb\\[2mm]
\phi_{47}^{(1)B}(\delta) &=& \f{34\, \delta^2 + 59\, \delta -18}{486}~ \f{\delta^2 \ln\delta}{1-\delta}  
                             ~+~ \f{433\, \delta^3+ 429\, \delta^2 -720\,\delta}{2916}.\label{phi47B}
\eea
The latter function is a new result from Ref.~\cite{Huber:2014nna} that originates from $s q
\bar q \gamma$ final states ($q=u,d,s$). Contributions to $b \to X_s^p\gamma$
from such final states at the NLO have been neglected in the previous
literature because they are suppressed by phase space factors
and the small Wilson coefficients $C_{3,\ldots,6}$.

\newappendix{Appendix~D:~~ Input parameters}
\def\theequation{D.\arabic{equation}}

In this appendix, we collect numerical values of the parameters that matter
for our branching ratio calculation in Section~\ref{sec:pheno}. The photon
energy cut is set to $E_0 = 1.6\,{\rm GeV}$. Our central values for the
renormalization scales are $\mu_b = \mu_c = 2.0\;{\rm GeV}$ and $\mu_0 = 160\;{\rm GeV}$. 

Masses of the $b$ and $c$ quarks together with the semileptonic $B\to
X_c\ell\bar\nu$ branching ratio ${\mathcal B}_{c\ell\bar\nu}$
and several non-perturbative parameters are
adopted from the very recent analysis in Ref.~\cite{Alberti:2014yda}.\footnote{
See also the previous version~\cite{Gambino:2013rza} where more details on the method are given.}
In that work, fits to the measured semileptonic decay spectra have been
performed with optional inclusion of constraints from the $b$-hadron
spectroscopy, as well as from the quark mass determinations utilizing moments of $R(e^+
e^- \to {\rm hadrons})$~\cite{Chetyrkin:2009fv}.  While $m_c$ is
$\overline{\rm MS}$-renormalized, $m_b$ and the non-perturbative parameters
are treated in the kinetic scheme. We choose the option where both $m_b$ and $m_c$
are constrained by $R(e^+ e^- \to {\rm hadrons})$, and $m_c(2\,{\rm GeV})$ is used in the fit. 
Once the parameters are ordered as~ $\{ m_{b,\rm kin},~ m_c(2\,{\rm
GeV}),~ \mu^2_\pi,~ \rho^3_D,~ \mu^2_G,~ \rho^3_{LS},~ 
{\mathcal B}_{c\ell\bar\nu}\}$ (expressed in GeV raised to appropriate powers),
their central values $\vec{x}$, uncertainties
$\vec{\sigma}$, and the correlation matrix $\hat{R}$ read~\cite{Gambino:2014priv}
\bea
\vec{x} &=&~ \left( \begin{array}{rrrrrrr} 
~~\,4.564 & ~~1.087 & ~~\,0.470 & ~~\,0.171 & ~~0.309 & \,-0.135 & ~~10.67  \end{array} \right),\nnb\\
\vec{\sigma} &=&~ \left( \begin{array}{rrrrrrr} 
~~\,0.017 & ~~0.013 & ~~\,0.067 & ~~\,0.039 & ~~0.058 &  ~~\,0.095 & ~~~~0.16 \end{array} \right),\nnb\\
\hat{R} &=& \left( \begin{array}{rrrrrrr} 
 1.000 &  0.461 & -0.087 &  0.114 &  0.542 & -0.157 & -0.061 \\  
 0.461 &  1.000 & -0.002 & -0.020 & -0.125 &  0.036 &  0.029 \\ 
-0.087 & -0.002 &  1.000 &  0.724 & -0.024 &  0.049 &  0.153 \\
 0.114 & -0.020 &  0.724 &  1.000 & -0.101 & -0.135 &  0.076 \\ 
 0.542 & -0.125 & -0.024 & -0.101 &  1.000 & -0.011 & -0.009 \\ 
-0.157 &  0.036 &  0.049 & -0.135 & -0.011 &  1.000 & -0.023 \\
-0.061 &  0.029 &  0.153 &  0.076 & -0.009 & -0.023 &  1.000 
\end{array} \right).\label{correl}
\eea 

Apart from the above parameters, the analysis of
Ref.~\cite{Alberti:2014yda} serves us as a source of a numerical formula
for the semileptonic phase-space factor
\be \label{phase}
C = \left| \f{V_{ub}}{V_{cb}} \right|^2 
\f{\Gamma[\bar{B} \to X_c e \bar{\nu}]}{\Gamma[\bar{B} \to X_u e \bar{\nu}]},
\ee
which reads~\cite{Gambino:2014priv}
\bea
C &=& g(z)\;  \left\{
0.903 - 0.588\, [\alpha_s(4.6\,{\rm GeV}) - 0.22] + 0.0650\, [m_{b,{\rm kin}} - 4.55]\right.\nnb\\[2mm] 
&-& \left. 0.1080\, [m_c(2\,{\rm GeV}) - 1.05]
- 0.0122\, \mu^2_G - 0.199\, \rho^3_D + 0.004\, \rho^3_{LS}\right\}, \label{cfit}
\eea
where $g(z) = 1 - 8 z + 8 z^3 - z^4 - 12 z^2 \ln z$~ and~ $z=
m^2_c(2\,{\rm GeV})/m^2_{b,{\rm kin}}$.~ Next, we use $C$ in the
expression~\cite{Gambino:2001ew}
\be \label{brB}
{\mathcal B}_{s\gamma}({E_{\gamma} > E_0})
= {\mathcal B}_{c\ell\bar\nu} 
\left| \f{ V^*_{ts} V_{tb}}{V_{cb}} \right|^2 
\f{6 \alpha_{\rm em}}{\pi\;C} 
\left[ P(E_0) + N(E_0) \right],
\ee
to determine the radiative branching ratio. Known contributions to the
non-perturbative correction $N(E_0)$ are given in terms of $\mu^2_\pi$,
$\rho^3_D$, $\mu^2_G$ and $\rho^3_{LS}$. The semileptonic branching ratio
${\mathcal B}_{c\ell\bar\nu}$ is CP- and isospin-averaged
analogously to Eq.~(\ref{brwidth}), while the isospin asymmetry effects in
both decay rates are negligible. Thus, neither the lifetimes nor the
production rates need to be considered among our inputs.

The remaining parameters that are necessary to determine $P(E_0)$ and the
overall factor in Eq.~(\ref{brB}) are as follows:
\mathindent0cm
\bea
\alpha_{\rm em}(0) &=&  1/137.036,\hspace{19mm}
M_Z ~=~ 91.1876\;{\rm GeV},\hspace{19.5mm}
M_W ~=~ 80.385\;{\rm GeV}~\mbox{\cite{Agashe:2014kda}},\nnb\\[2mm]
\alpha_s (M_Z) &=& 0.1185 \pm 0.0006~\mbox{\cite{Agashe:2014kda}},
\hspace{1cm} m_{t,{\rm pole}} ~=~ (173.21 \pm 0.51 \pm 0.71) 
\;{\rm GeV}~\mbox{\cite{Agashe:2014kda}},\nnb\\[2mm]
\left|\f{V_{ts}^* V_{tb}}{V_{cb}}\right|^2 &=& 0.9626 \pm 0.0012~\mbox{\cite{Charles:2015gya}},\hspace{21mm}
\f{m_b}{m_q} ~\in~ (10,50).\label{setpar1}
\eea
\mathindent1cm
For the electroweak and ${\mathcal O}(V_{ub})$ corrections to $P(E_0)$, we also need
\bea
\alpha_{\rm em}(M_Z) &=&  1/128.940,\hspace{3cm}
\sin^2\theta_W ~=~ 0.23126~\mbox{\cite{Agashe:2014kda}},\nnb\\[2mm]
M_{\rm Higgs} &=& 125.7 {\rm GeV}~\mbox{\cite{Agashe:2014kda}},\hspace{16mm}
\f{V_{us}^* V_{ub}}{V_{ts}^* V_{tb}} ~=~ -0.0080 + 0.018\, i~\mbox{\cite{Charles:2015gya}}.\label{setpar2}
\eea
The quark mass ratio $m_b/m_q$ ($q=u,d,s$) in Eq.~(\ref{setpar1}) serves as a
collinear regulator wherever necessary. Fortunately, the dominant
contributions to $\Gamma(b \to X_s^p\gamma)$ are IR-safe, while all the
quantities requiring such a collinear regulator contribute at a sub-percent
level only. They undergo suppression by various multiplicative factors
($C_{3,\ldots,6}$,~ $Q_d^2 \alpha_s/\pi$,~ etc.), and by phase-space
restrictions following from the relatively high $E_0 \sim m_b/3$. Changing
$m_b/m_q$ from 10 to 50 affects the branching ratio by around $0.7\%$ only. We
include this effect in our parametric uncertainty even though the dependence
on $m_b/m_q$ is spurious, i.e.\ it should cancel out once the non-perturbative
correction calculations are upgraded to take collinear photon emission into
account (see Refs.~\cite{Ferroglia:2010xe,Kapustin:1995fk,Asatrian:2013raa}). Thus, the
parametric uncertainty due to $m_b/m_q$ might alternatively be absorbed into
the overall $\pm 5\%$ non-perturbative error~\cite{Benzke:2010js}. Our
range for $m_b/m_q$ roughly corresponds to the range $[m_B/m_K, m_B/m_\pi]$,
which is motivated by the fact that light hadron masses are the physical
collinear regulators in our case.

All the uncertainties except for those in Eq.~(\ref{correl}) are treated as
uncorrelated. One should remember though that the dependence of $C$ on $\al$
is taken into account via Eq.~(\ref{cfit}).

\end{document}